\documentclass[12pt]{article}

\usepackage{scicite}

\usepackage{times}

\usepackage{color}
\usepackage{setspace}
\usepackage{graphicx}
\usepackage{subfigure}
\usepackage[percent]{overpic}
\usepackage{graphics}
\usepackage{caption}
\usepackage{adjustbox}
\usepackage{epsfig}
\usepackage{epstopdf}
\usepackage{amsmath}
\usepackage{bm}
\usepackage{amssymb}
\usepackage{booktabs}
\usepackage[table,xcdraw]{xcolor}
\usepackage{multirow}
\usepackage{makecell}
\usepackage{tabularx}
\usepackage{chngcntr}
\usepackage{enumitem}
\usepackage{hyperref}

\newenvironment{sciabstract}{%
\begin{quote} \bf}
{\end{quote}}

\title{On the Expressive Power of Behavior Structure}

\author
{Cheng Wang$^{1,2,3\ast}$, Hangyu Zhu$^{1,2}$, Yuhang Lin$^{1,2}$,  Changjun Jiang$^{1,2,3}$ \\
\\
\normalsize{$^{1}$Department of Computer Science and Technology, Tongji University;}\\
\normalsize{Shanghai, 201804, China}\\
\normalsize{$^{2}$Key Laboratory of Embedded System and Service Computing, Ministry of Education;}\\
\normalsize{Shanghai, 201804, China}\\
\normalsize{$^{3}$Shanghai Artificial Intelligence Laboratory; Shanghai, 200232, China}\\
\\
\normalsize{$^\ast$Corresponding author. Email: cwang@tongji.edu.cn}
}

\date{}

\begin{document}
\begin{spacing}{1.0}

\baselineskip24pt


\maketitle


\begin{sciabstract}

Efforts toward a comprehensive description of behavior have indeed facilitated the development of representation-based approaches that utilize deep learning to capture behavioral information. As behavior complexity increases, the expressive power of these models reaches a bottleneck. We coin the term ``behavioral molecular structure" and propose a new model called the Behavioral Molecular Structure (BMS). The model characterizes behaviors at the atomic level, analogizes behavioral attributes to atoms, and concretizes interrelations at the granularity of atoms using graphs. Here, we design three different downstream tasks to test the performance of the BMS model on public datasets. Additionally, we provide a preliminary theoretical analysis demonstrating that the BMS can offer effective expressiveness for complex behaviors.
\end{sciabstract}


\clearpage

\paragraph*{Behavioral molecular structure awakens a profound behavior analysis}

\paragraph*{}
Efforts have been made to express behavior accurately, allowing for quantitative representations of complex behaviors with significant improvements, but there are limitations to such representations. Drawing inspiration from the interactions between atoms in the material world that form molecules, Wang et al. coin the term behavioral molecular structure and design a novel general solution for expressing behavior. They imagine that behavior can be concretized as a molecular structure that reflects the interrelationships between its attributes. The results on various real-world behavioral datasets suggest that, compared to behavioral representations, behavioral structures can further improve the expression of behavior.


\section*{Introduction}

A deep understanding of the movement in the objective world opens up a new pathway for behavioral expression.
Behavior refers to any observable and measurable action, response, or activity \cite{mathis2020deep}.
When we observe and measure behavior, there is often a significant gap between our expectations of its expressive power and the current behavior modeling paradigm's ability to capture behavioral expression \cite{gallego2017neural}.
A complete and comprehensive description of behavior will require large amounts of data, such as high-resolution video clips, to capture the full range of actions, responses, and activities involved \cite{branson2009high}.
Processing and analyzing such behavior data is undoubtedly computationally intensive \cite{pereira2019fast}.
Reducing such behavioral descriptions into more expressive observational features has significant practical importance \cite{sanger2000human}.
Going even further, it is a triumph that simplifies behavioral descriptions into low-dimensional representations for just limited degrees of freedom using representation-based methods \cite{bialek2022dimensionality,mathis2018deeplabcut,yu2008gaussian}.

Our research is driven by the observation that, in the material world, molecules have structures that can be described by chemical bonds representing strong interactions between neighboring atoms (or ions) and fine-grained atoms \cite{chen2017ab,bartok2017machine,hansen2015machine}.
Similarly, we have reason to believe that behavior, as an observable objective reality, has a measurable structure that can be explored and analyzed using a variety of tools and techniques.
Basic scientific research (e.g., mathematical hypotheses used in physics) on the material world can provide insights into the behavior under complex conditions.
By uncovering this underlying mechanism (like structure), researchers can gain new insights into the mechanisms that govern behavior, leading to more accurate and sophisticated models of behavior. This, in turn, can help to inform the development of more effective behavior models for a wide range of subtasks.

From human social activities to animal movements in the biological world, to air flow in the natural world, all descriptions of behavior are subject to limitations of predictability and diversity \cite{teixeira2019deciphering,kulshrestha2021web,lipinski2016cancer,guilyardi2012new,colizza2006role}.
When we talk about behavior, we usually assume that individual behavior is not entirely random and is predictable to some extent \cite{doi:10.1126/science.1177170,smolak2021impact,chen2022contrasting,zhang2022beyond}.
However, from the perspective of an external observer, the behavior seems to be relatively random (i.e. there are differences between behaviors) to the behavior of others \cite{doi:10.1126/science.1177170,castellano2009statistical,brockmann2006scaling,gonzalez2008understanding,colizza2007modeling,hufnagel2004forecast}.
So it is possible to develop high-expressive models of behavior under certain conditions, such as detecting malicious behavior \cite{ali2019detect,pantelaios2020you,he2021datingsec}, predicting behavior trajectories \cite{phan2020covernet,ngiam2021scene,mozaffari2020deep}, generating simulated behavior \cite{feng2021intelligent,chen2021decision,schwarting2019social}, etc.

Behavior can have a multi-level description structure that is able to encode information.
Various methods can be employed to provide high-precision observational data for describing behavior.
After decades of research, representation-based methods, utilizing the nonlinear transformation ability of deep learning, have pushed the understanding of behavior to a higher level \cite{nestor2016feature,hsu2021b,yan2020unexpected}. Behavioral representations have greater expressive power than direct observation of behavior. From the beginning of association within behavior sequences to association in behavior communities, increasingly complex behavior models have been developed. They quantified the regularity of behavior using representations and clustered or separated them in low-dimensional space. Unfortunately, as the complexity of behavior increases, the expressive power of these models reaches a bottleneck. Further improvement in behavioral expression is undoubtedly a persistent challenge in the scientific community.

Understanding behavior, especially studying the correlation between behaviors, is a sophisticated issue. Facing two behaviors, one qualitative answer is to compare and discover their differences and predictability by their consistent/inconsistent.
Just as in genetics, we analyze the DNA structure to achieve a refined quantitative analysis (which can be seen as a phenomenological description of genetic patterns) to explain why we resemble our parents or why we differ in appearance from others.
In broad contexts, the structure of behavior is a candidate for a unified behavioral description. How can we construct a complex structure of behavior to further enhance our ability to express behavior?
All of these puzzles point to the problem of defining the behavioral structure.
In mathematics, the structure of a mathematical object refers to a non-empty basic set $A$ (the research object) and a set of relationships $R$ that satisfy certain properties defined on the basic set $A$ \cite{shurman2016calculus,aluffi2021algebra,szekeres2004course,lynn2020humans}.

Taking inspiration from the interactions between atoms in molecular structures \cite{li2023reaction}, we simulate internal correlations in behavior.
Because this phenomenon is not widely known, we coin the term \emph{behavioral molecular structure} to facilitate research in this field.

To work toward a sharper definition, consider the case in which the behavior is limited behavioral attributes, all the behavioral attributes consist of the basic set $A$. Furthermore, we obtain the set of relationships $R$ after determining the relationships between behavioral attributes.
Thus, we can concretize observable behavior as a structure defined on sets $A$ and $R$ (Fig. \ref{Fig1}A).

Identifying key structural features is crucial for revealing molecular activities and properties, and the same applies to behavior.
The behavioral molecular structure should capture the characteristics and properties of various behaviors. An excellent behavioral molecular structure should meet the following conditions:

$\bullet$ Uniqueness.
Each behavior has its unique structure and properties. Even very similar behaviors may have different structures and properties due to small differences.

$\bullet$ Stability.
Behaviors usually maintain their key structure and properties under certain conditions (such as time and place).
Despite the apparent randomness of individual behavior, the daily behavior patterns hide unexpectedly high potential predictability.

The behavioral space is close to infinity, but it can be limited by decomposing behavior into a combination of finite behavioral attributes (just like atoms in molecular structure). In this way, the behavioral expression is transformed into different combinations of behavioral attributes. A behavioral attribute (like a single atom) is influenced by other behaviors, and perturbations to the behavioral molecular structure can lead to huge differences between behaviors.

To demonstrate the effectiveness of behavioral structure, we argue for the rationality of behavior molecular structure by examining its expressive power through theoretical analysis. Furthermore, we validate its effectiveness by deploying it in tasks that involve fundamental behavioral functions (such as detection, prediction, and generation).
Specifically, we introduce a general solution, the behavioral molecular structure (BMS) model. BMS characterizes behaviors at the atomic level, with the behavioral attributes that make up the behavior being likened to atoms, where the interrelationships between attributes are studied and concretized using graphs.
Given a candidate space consisting of potential behavioral attributes, the behavioral molecular structure model can reconstruct behavior as a graph consisting of behavioral attributes and their interactions, based on the interplay between behavioral attributes (Fig. \ref{Fig1}B). By allocating and learning the structure and features of the graph, we introduce an interface for further processing of behavior structure within existing neural network architectures.

\subsection*{Behavioral expressive power}

In deep learning, the expressive power of a model can be estimated by the count of linear regions that are partitioned by the model \cite{hanin2019complexity}.
Models with more linear regions tend to better capture relationships within the data.
Similarly, we introduce the concept of measuring behavioral expressive power by considering the number of distinct behaviors into which models partition the behavioral space.
In this context, a behavior is a set of features, referred to as behavioral attributes, that are used for behavioral description.

\begin{figure*}[!tbh]
    \centering
	\begin{overpic}[width=0.225\textwidth]{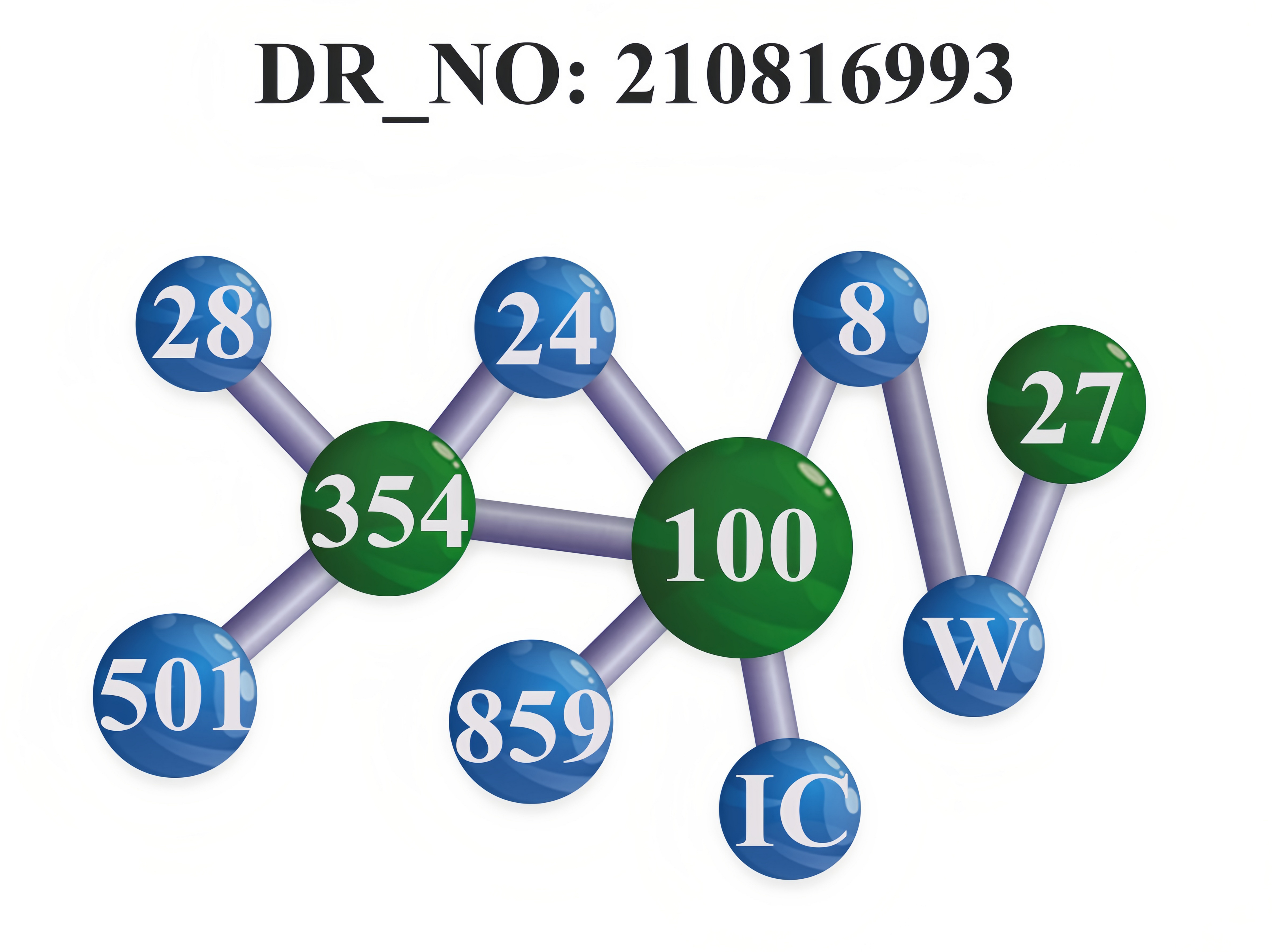}\label{figure:structure_toy}
			\put(-0,80){\large\textbf{A}}
		\end{overpic}
		\hspace{2mm}
		\begin{overpic}[width=0.225\textwidth]{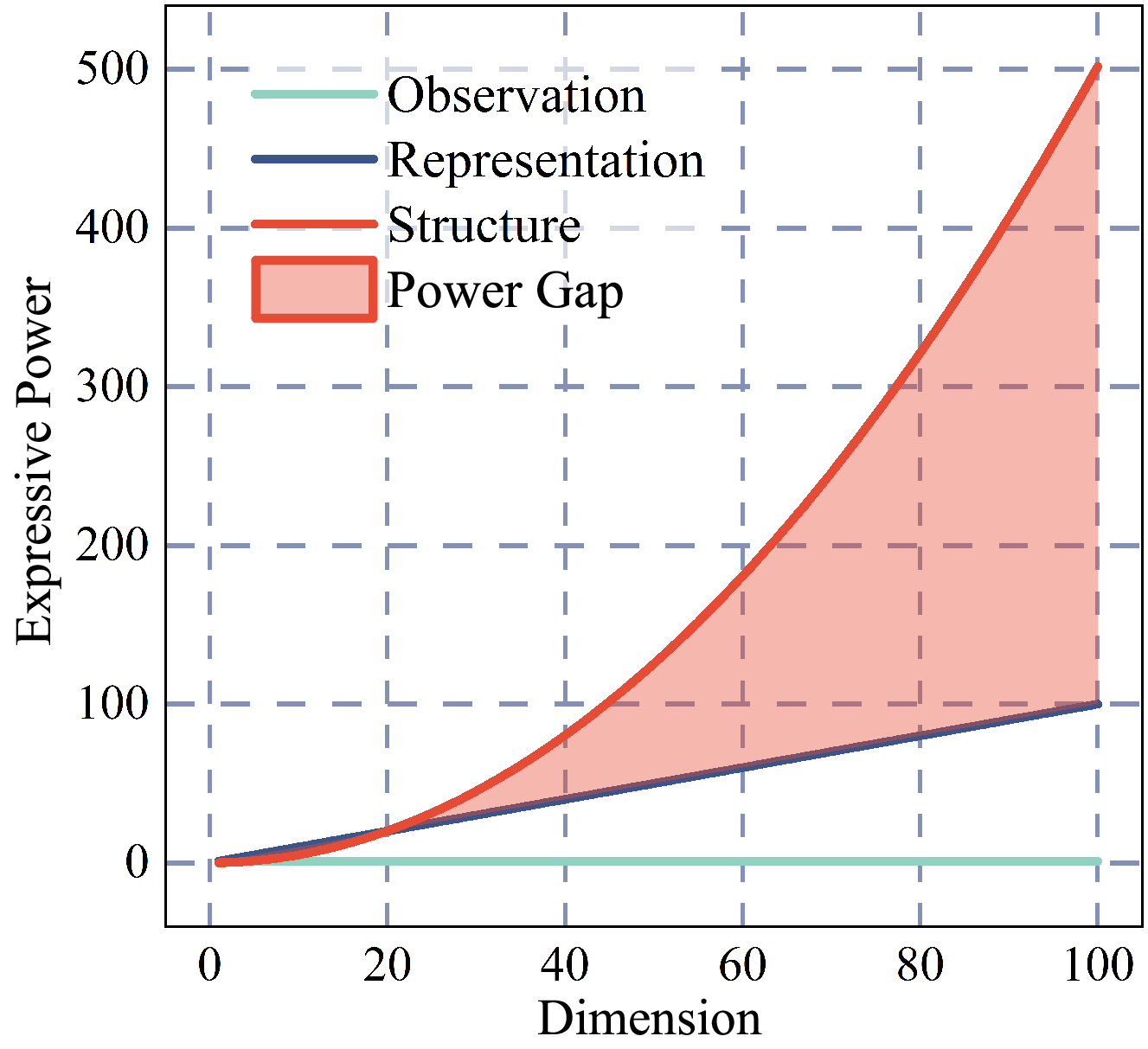}\label{figure:expressive_power}
			\put(-21,80){\large\textbf{C}}
		\end{overpic}
		\hspace{2mm}
		\begin{overpic}[width=0.225\textwidth]{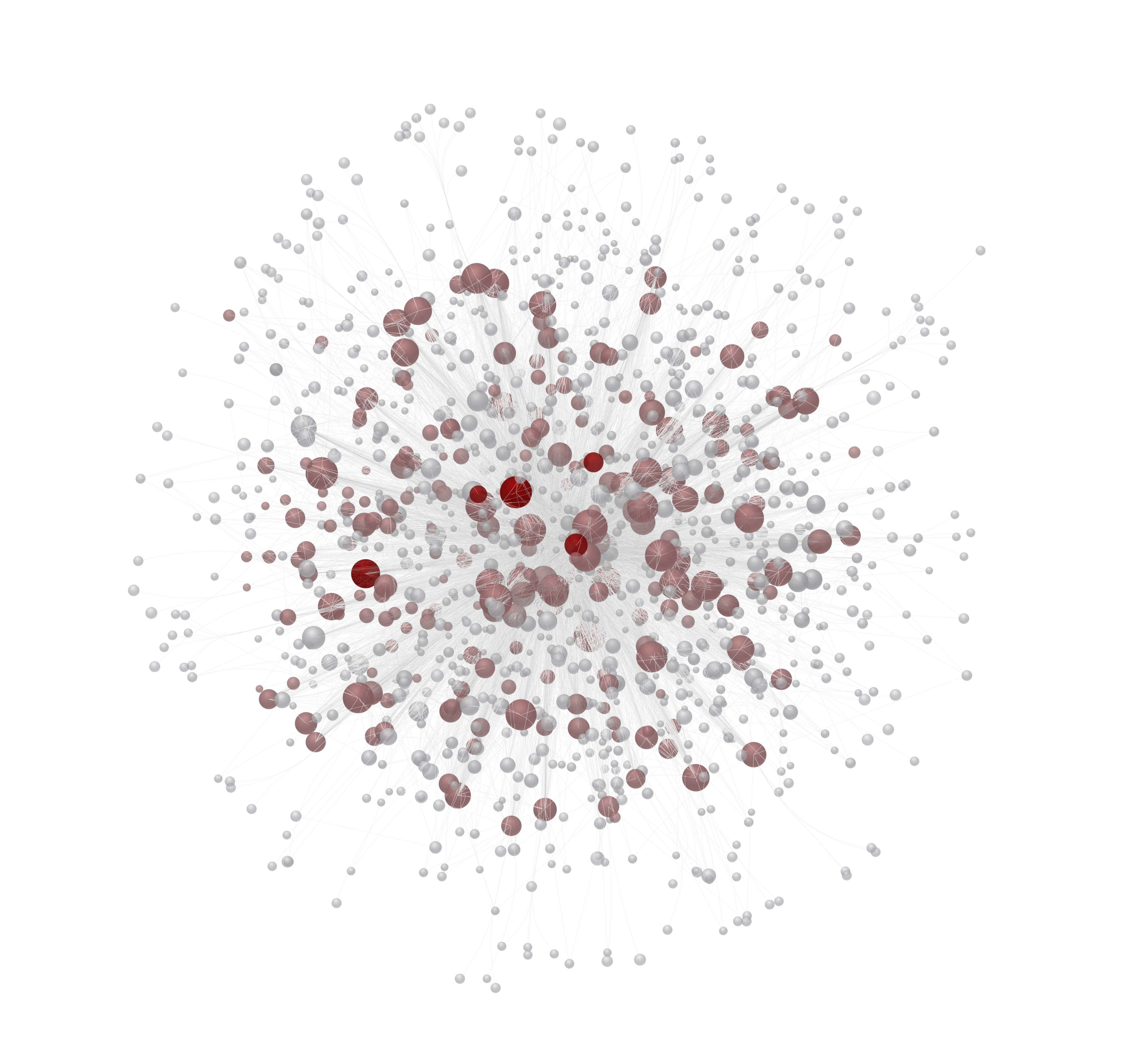}\label{figure:expressive_power}
			\put(-1,80){\large\textbf{D}}
		\end{overpic}
		\hspace{2mm}
		\begin{overpic}[width=0.225\textwidth]{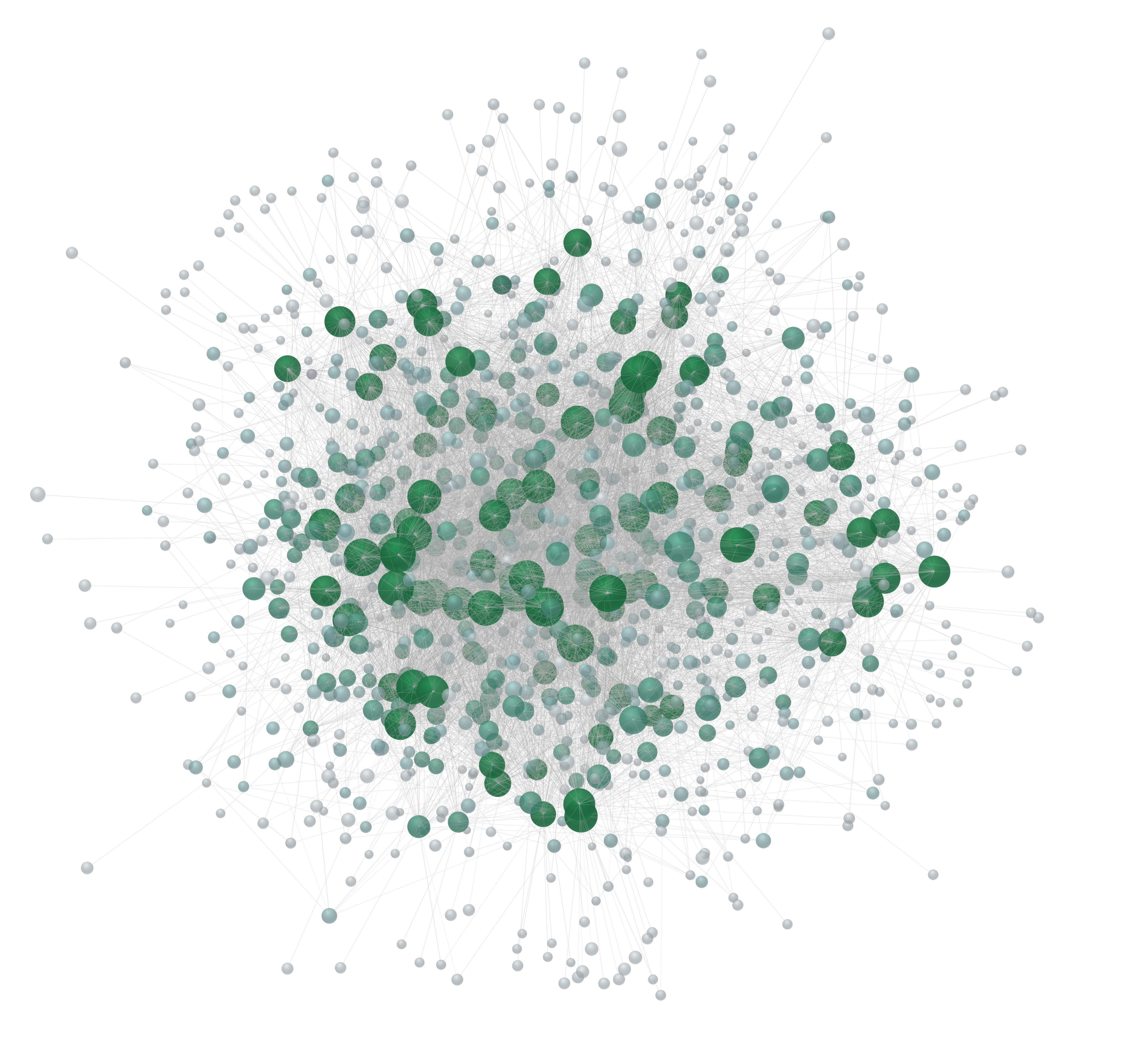}\label{figure:expressive_power}
			\put(-1,80){\large\textbf{E}}
		\end{overpic}

		\begin{overpic}[width=0.995\textwidth]{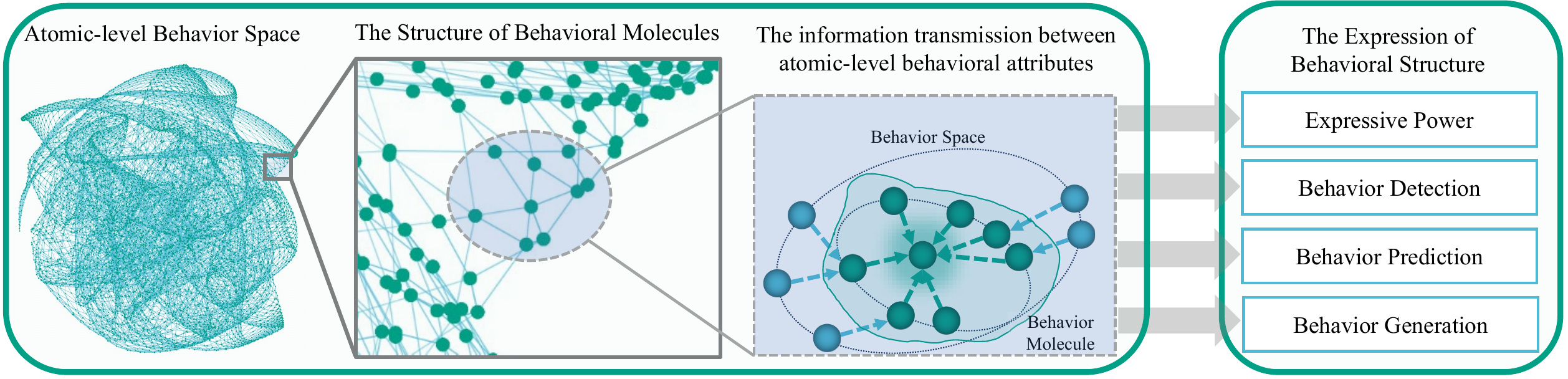}\label{figure:bms}
			\put(-0,24){\large\textbf{B}}
		\end{overpic}

		\captionsetup[subfigure]{labelformat=empty}
		\caption{\textbf{Behavioral molecular structure and the demonstration of its expressive power.}
(\textbf{A})  Illustration of behavioral molecular structure. The behavior is visualized as a structure composed of nodes and edges, where we select ten behavior attributes for the criminal behavior (numbered 210816993).
(\textbf{B}) BMS takes as input a complete behavioral attribute space, specified by all potential behavioral attributes and molecular structures defined on each behavior. The embeddings based on atomic-level behavioral attributes within the behavior molecule and in the contextual environment are repeatedly updated. This process generates a set of embeddings that encode each behavioral attribute. Then, graph pooling is performed on each behavior molecule, and the pooled embeddings are input into additional downstream neural network layers. Finally, we perform downstream tasks (such as detection, prediction, and generation) based on the pooled results obtained from the behavioral molecule structure.
(\textbf{C}) Comparison of expressive power of different behavioral expressions on behavioral dimensions. We implement logarithmic operation on the value of expressive power to facilitate display.
(\textbf{D})   Use a visualization algorithm to perform visualization of the structure of 2000 behaviors each consisting of 10 behavioral attributes, and randomly select one behavior to colorize based on the neighbor of each of its behavioral attributes. The resulting graph provides an example of how behaviors are positioned in behavioral space and how they influence contextual behaviors.
(\textbf{E})  Behavioral attribute frequency associated with these behaviors shown in (D). The area of a node corresponds to the number of times the corresponding behavioral attribute appears in the behavioral data, and the degree of a node is roughly proportional to its area. Behavioral attributes that appear frequently tend to be located in key positions in the behavioral structure.
}

		\label{Fig1}
	\end{figure*}

Fig. \ref{Fig1}C illustrates the behavioral expressive power of three representative models: observation-based, representation-based, and structure-based models.
Observation-based models exhibit a constant level of behavioral expressive power as they are irrespective of behavioral dimension.
Representation-based models provide an exponential level of behavioral expressive power.
They can divide the behavioral space into the number of behavioral value combinations.
Given behavioral data of dimension $n$, where the number of values for the behavioral attribute on the $i$th dimension is denoted as $k_i$, the number of all possible behaviors can be calculated as $\prod_{1}^{n}k_i$. By scaling $k_i$, taking $k:=\max(k_1,\cdots,k_n)$, the theoretical upper limit of behavioral expressive power for such models is $k^n$ ($k$ is set to $100$ in Fig. \ref{Fig1}C).
Structure-based models enable the translation of behavioral expression into nodes' connectivity in the graph. Given behavioral data of dimension $n$, these models can express $k^{n\cdot(n-1)}$ behaviors. Here, $k$ represents the possible states of connectivity between nodes, such as connected or unconnected, with $k=2$ in Fig. \ref{Fig1}C. Hence, the structure-based models can theoretically surpass ordinary exponential level $k^n$.
It is crucial to emphasize that possessing a high level of expressive power does not guarantee superior performance and should not be considered the sole indicator of high performance.

The notable result is that with the increase of behavioral dimension, behavioral structure has a significant advantage compared to the other two ways of expression. We also notice that when the behavioral dimension is less than $20$, the expressive power of behavioral structure is slightly lower than that of behavioral representation. This shows that when behaviors only require relatively few features to describe, facing such a simple behavior, behavioral representation can sufficiently describe the behavior.

Representation-based models can use the similarity between representations to compare the semantic information of two behaviors.
We use a similar method to estimate the semantic influence between behaviors in structure-based behavioral research. To illustrate this, we visualize $2,000$ criminal behaviors from an open dataset. To ignore the effects of structure selection, we assume that there is the correlation between any two behavioral attributes within each behavior. We study the influence of behavioral attributes on contextual behaviors and their impact on behavioral semantics. In Fig. \ref{Fig1}D, we find that a behavior's semantic information can have an impact on almost the entire behavioral attribute space through indirect correlations between behavioral attributes. Fig. \ref{Fig1}E shows that frequently occurring behavioral attributes tend to be located in key positions within behavioral attribute communities, with more connections transmitting their information to other behaviors.

\subsection*{Structure is what you need}

We deploy BMS to identify different individuals' crime record categories in the city of Los Angeles since 2020 (transcribed from the original printed crime reports). The raw data contains over $700,000$ criminal incidents. Here we focus on identifying the $10$ most common types of criminal incidents.

\begin{figure*}[!t]
	\begin{overpic}[height=0.255\textwidth]{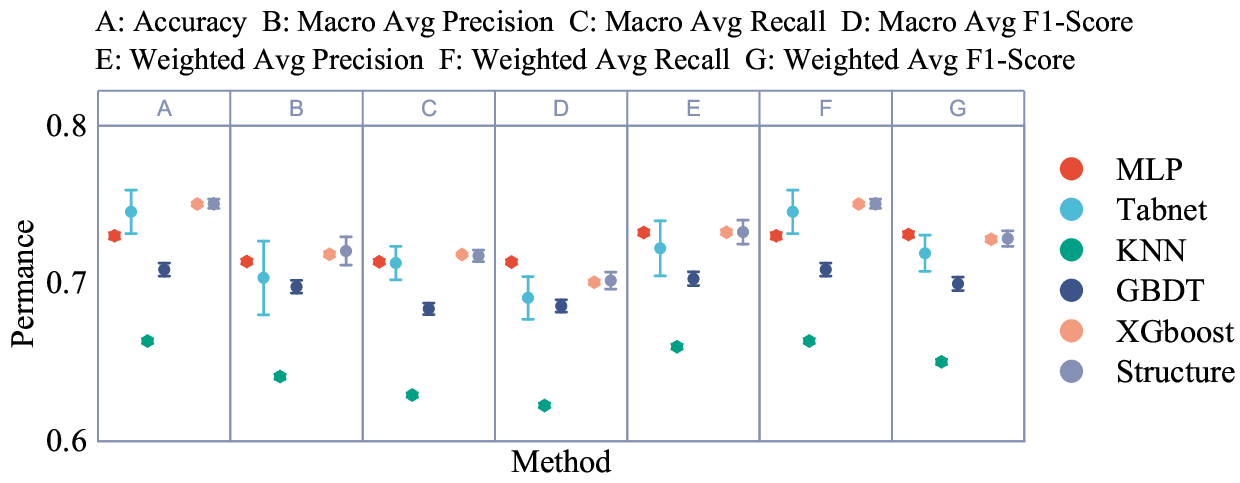}
			\put(-1,34.6){\large\textbf{A}}
		\end{overpic}
		\hspace{2mm}
		\begin{overpic}[height=0.26\textwidth]{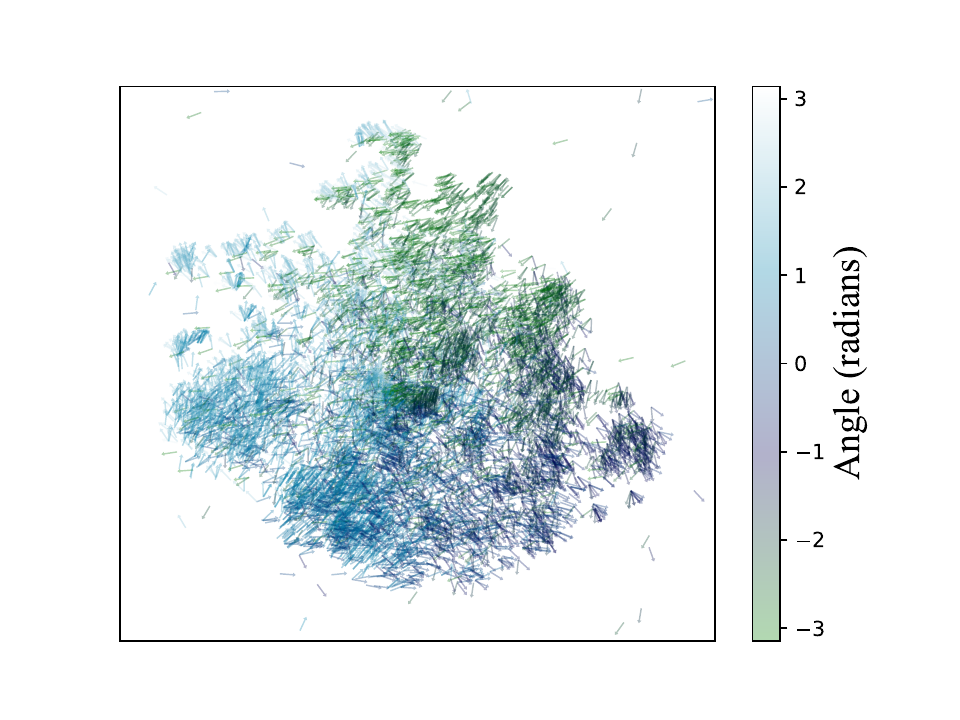}
			\put(-5,66){\large\textbf{B}}
		\end{overpic}

		\begin{overpic}[height=0.235\textwidth]{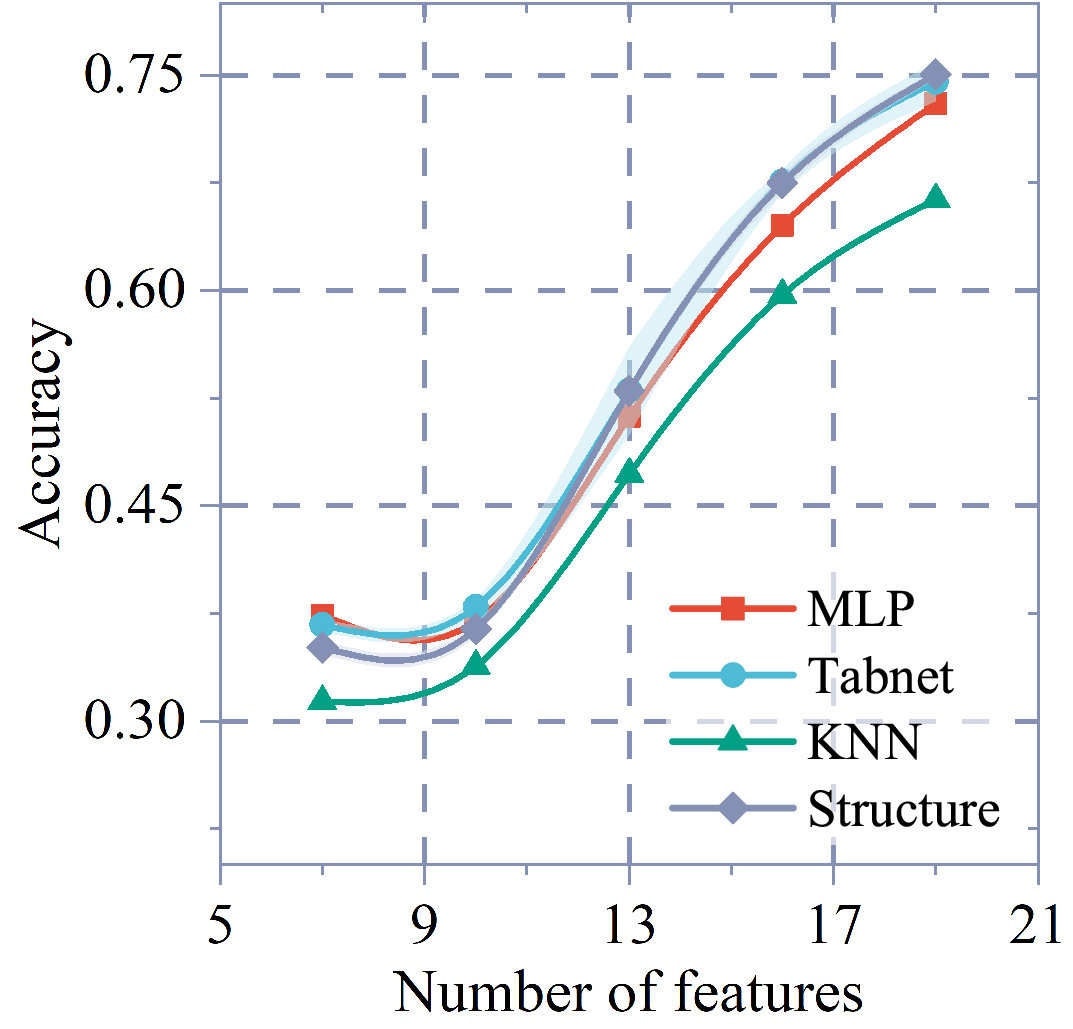}
			\put(-4,87.9){\large\textbf{C}}
		\end{overpic}
		\hspace{0.5mm}
		\begin{overpic}[height=0.235\textwidth]{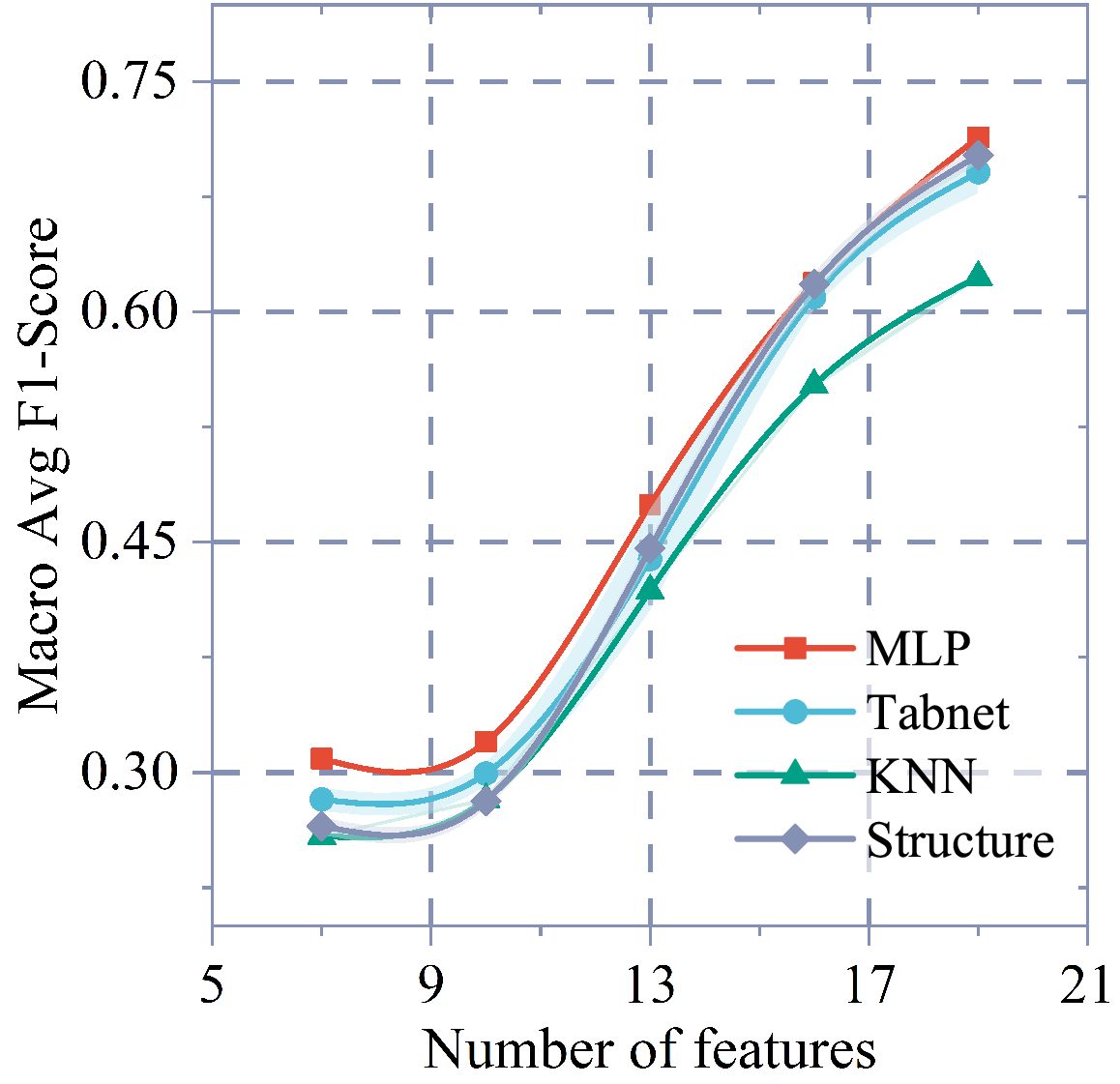}
			\put(-4,88){\large\textbf{D}}
		\end{overpic}
		\hspace{1mm}
		\begin{overpic}[height=0.235\textwidth]{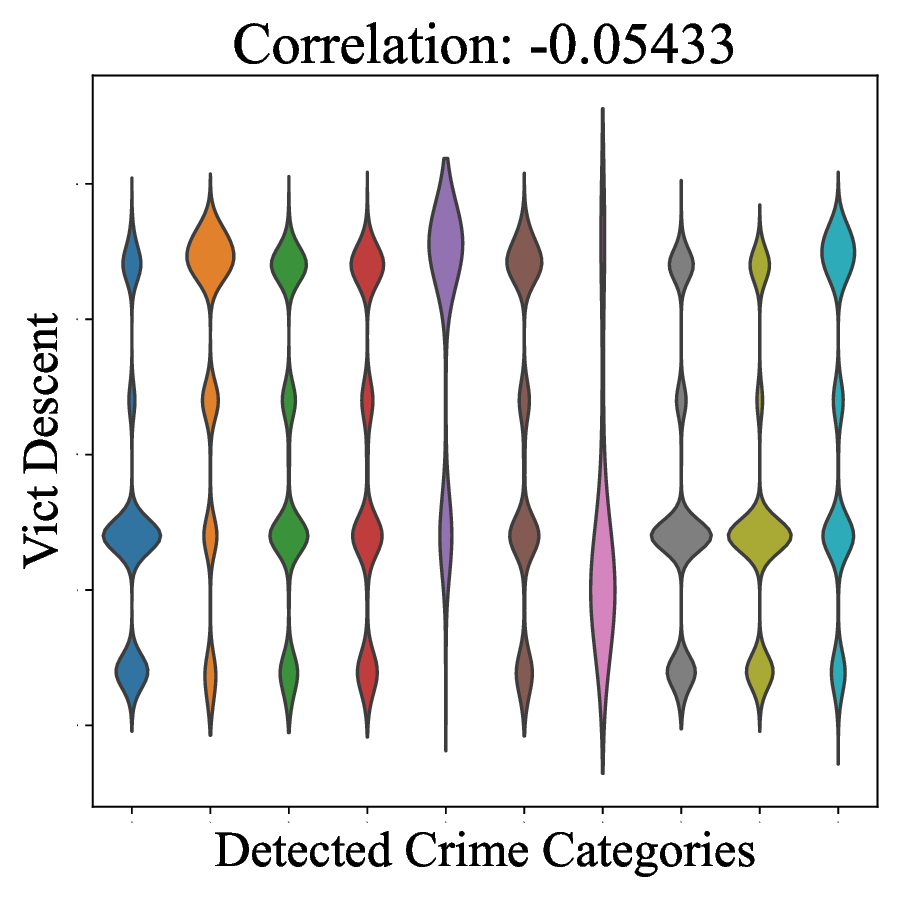}
			\put(-4,91){\large\textbf{E}}
		\end{overpic}
	   \hspace{1mm}
		\begin{overpic}[height=0.235\textwidth]{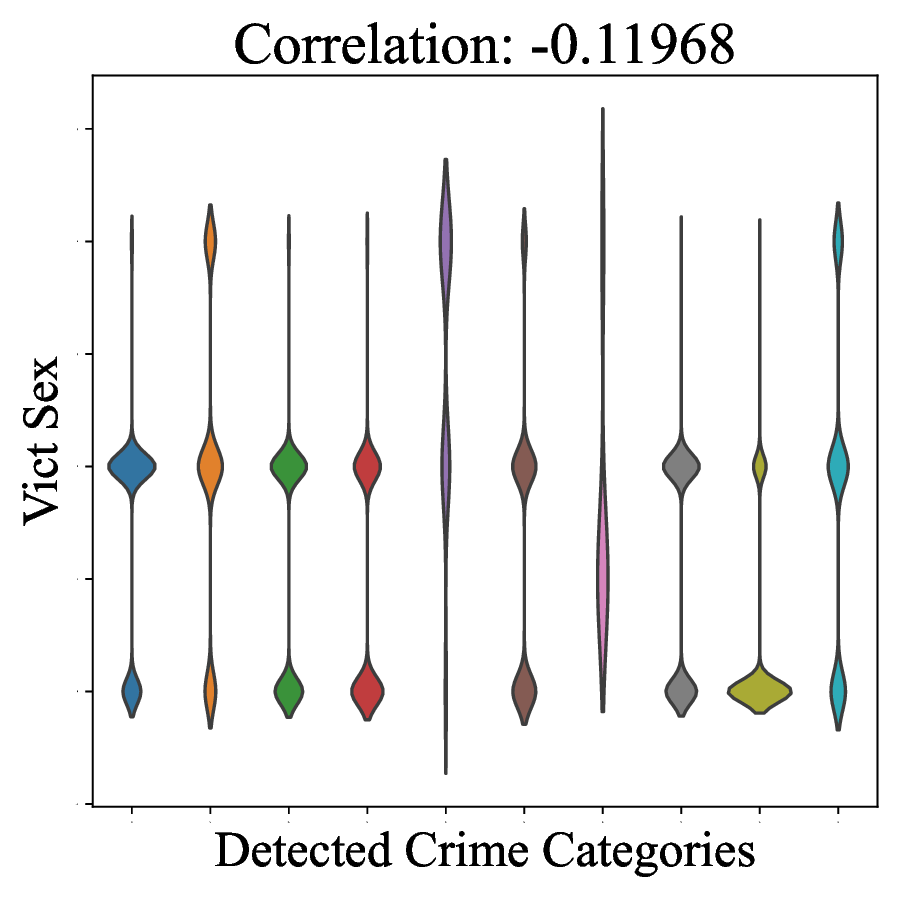}
			\put(-4,91){\large\textbf{F}}
		\end{overpic}
		\begin{overpic}[height=0.1885\textwidth]{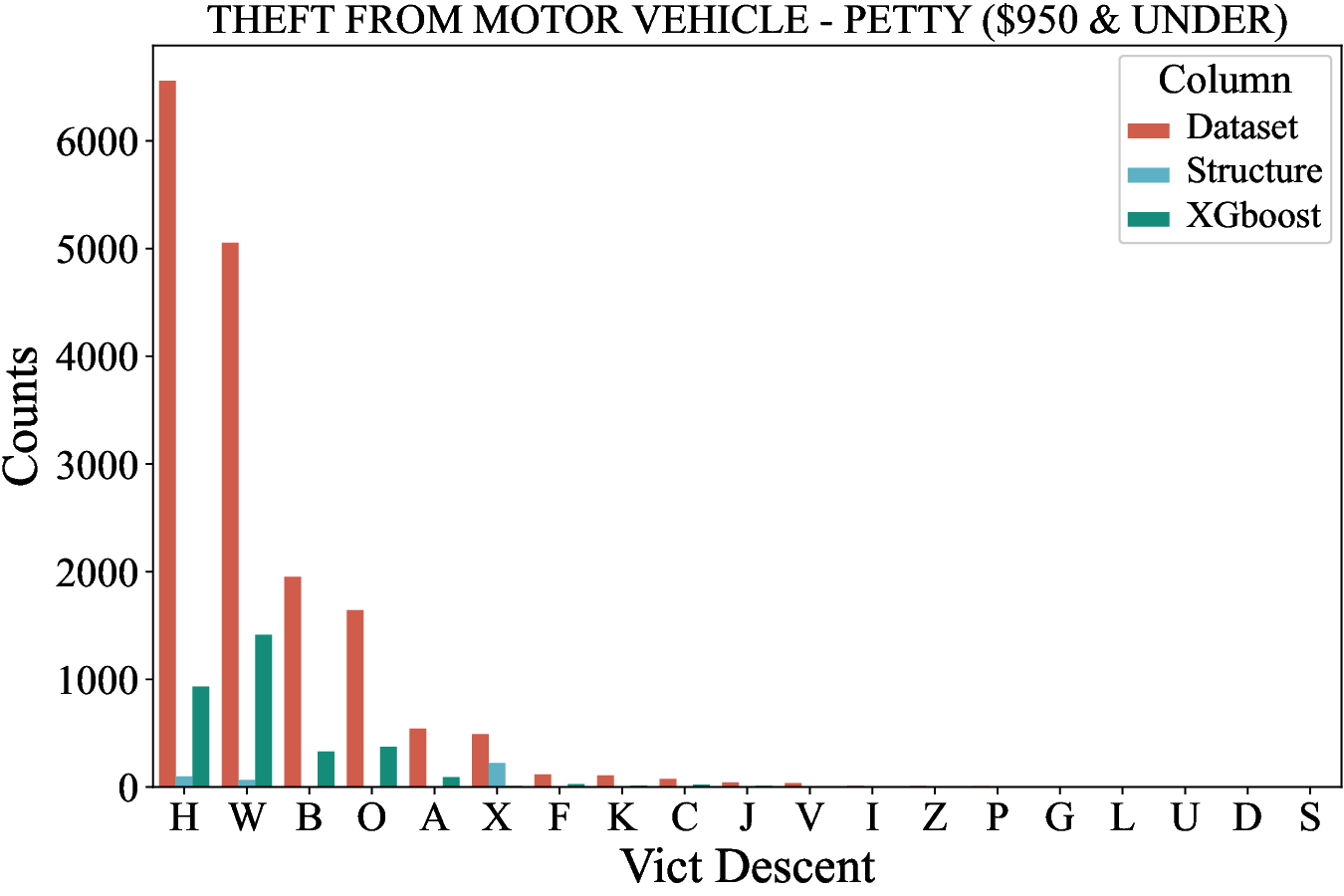}
			\put(-4,58){\large\textbf{G}}
		\end{overpic}
		\hspace{1mm}
		\begin{overpic}[height=0.1885\textwidth]{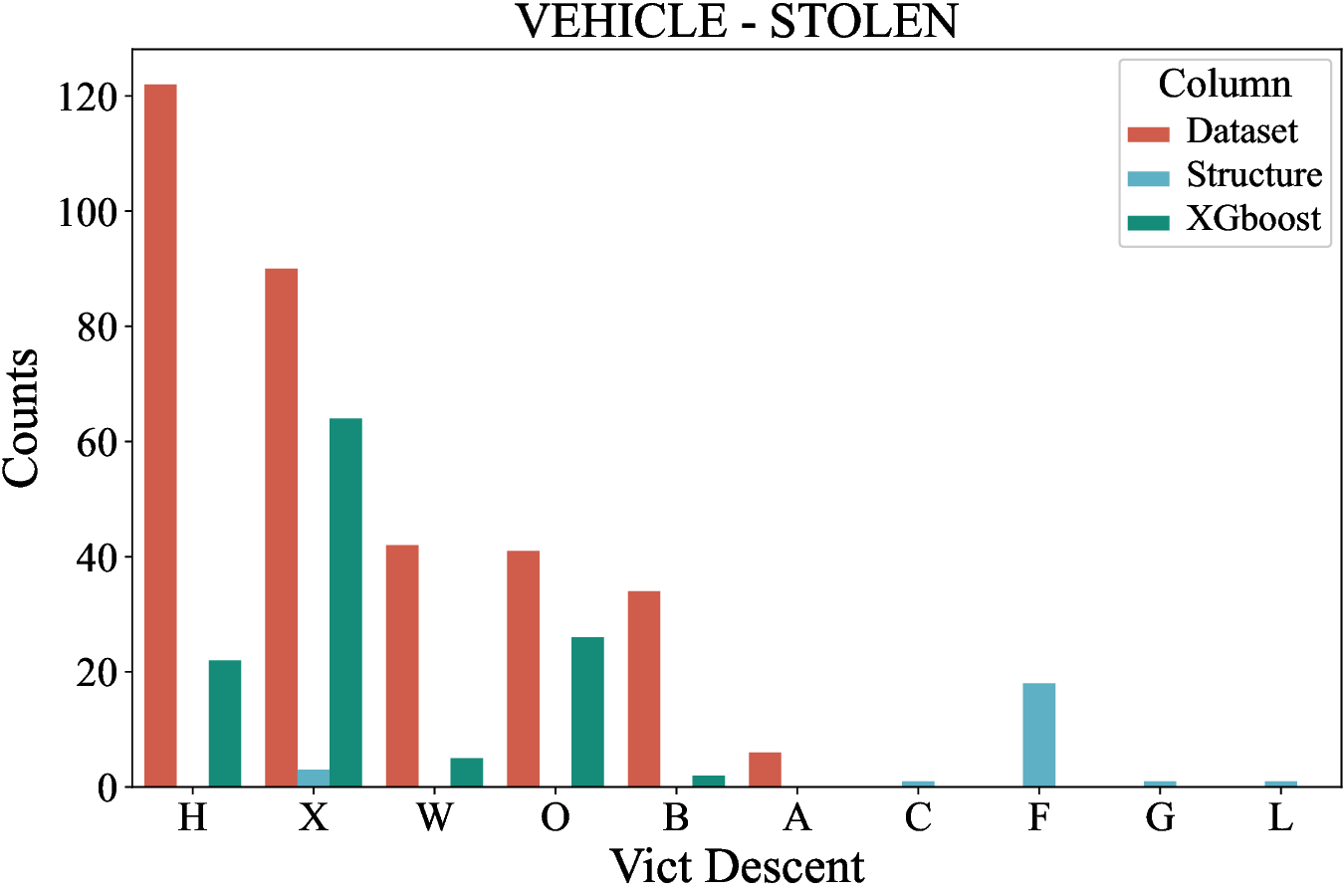}
			\put(-6,58){\large\textbf{H}}
		\end{overpic}
		\hspace{1mm}
		\begin{overpic}[height=0.1885\textwidth]{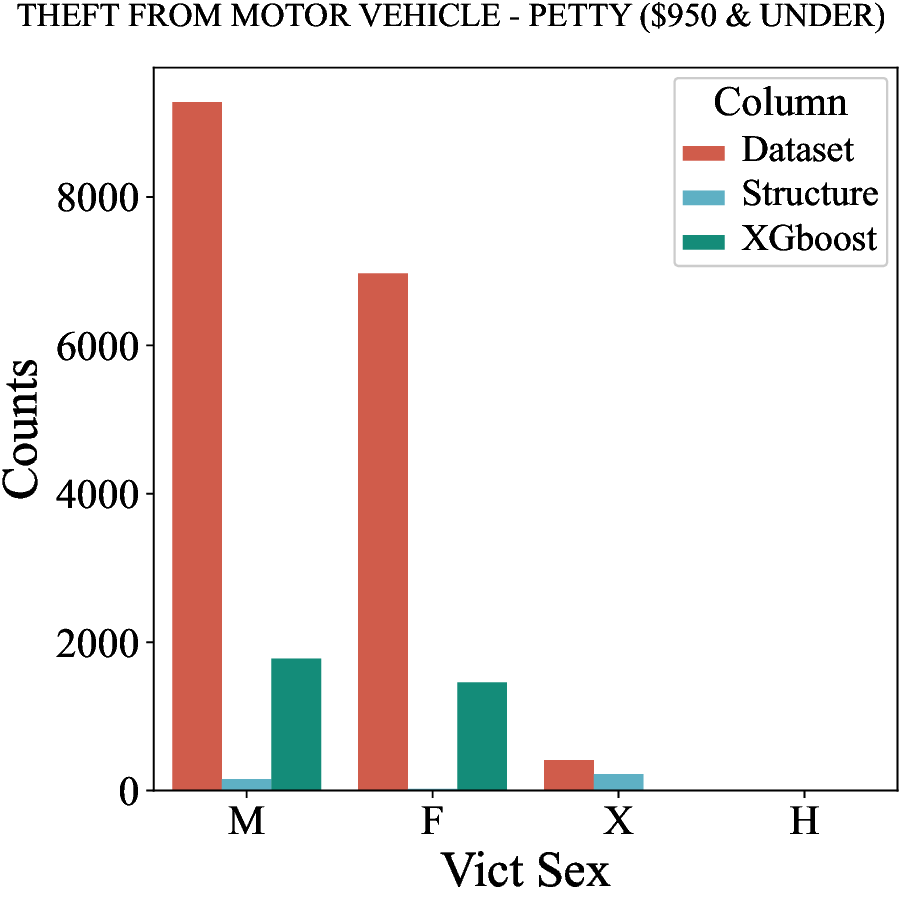}
			\put(-6,88){\large\textbf{I}}
		\end{overpic}
		\hspace{1mm}
		\begin{overpic}[height=0.1885\textwidth]{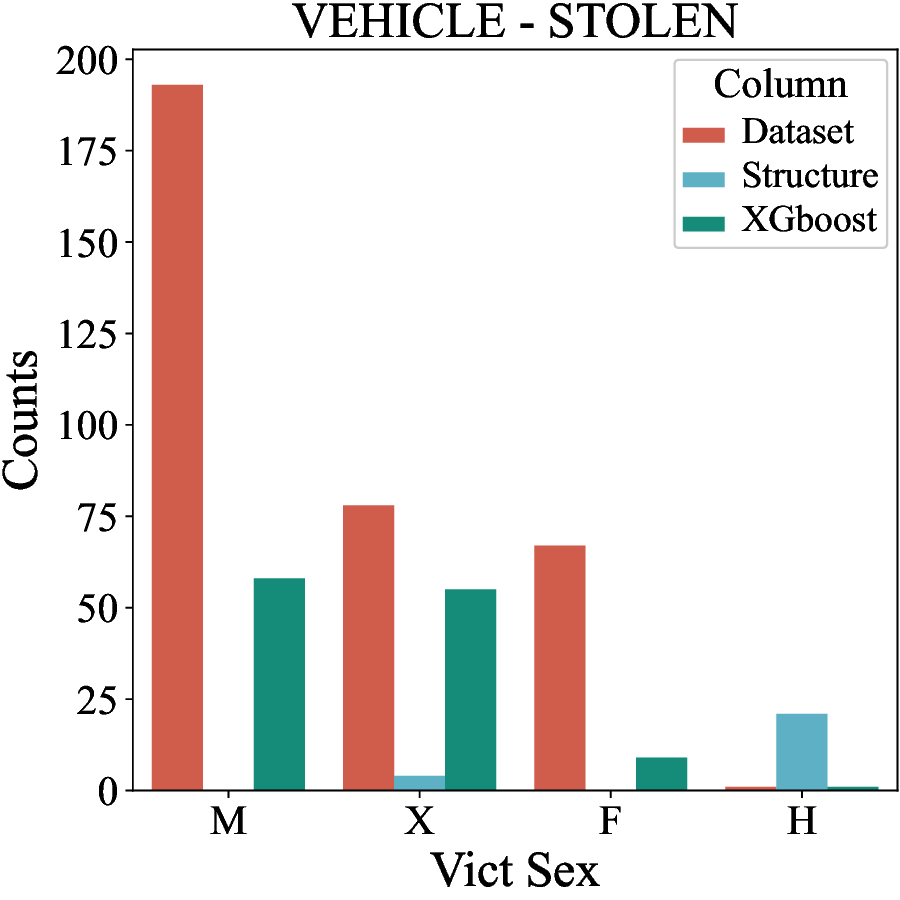}
			\put(-6,88){\large\textbf{J}}
		\end{overpic}

		\captionsetup[subfigure]{labelformat=empty}
		\caption{\textbf{Superior performance on behavioral detection.}
(\textbf{A}) In terms of identifying the category of criminal behaviors, BMS surpasses typical methods by utilizing behavioral molecular structures on multiple representative classification metrics.
(\textbf{B}) The movement trends between the vectors obtained from behavior attributes after representation learning based on behavioral molecular structures and the initial semantic vectors. Lighter colors indicate smaller angles between the vectors.
(\textbf{C}) The trend of accuracy in behavior detection as  the number of behavior attributes changes.
(\textbf{D}) The trend of macro average F1-score in behavior detection as the number of behavior attributes changes.
(\textbf{E}) The correlation between the crime behavior categories predicted by BMS and the ethnicity of the victims.
(\textbf{F}) The correlation between the crime behavior categories predicted by BMS and the gender of the victims.
(\textbf{G})-(\textbf{J}) display the statistical data regarding the identification of criminal behaviors (``theft from motor vehicle-petty (\$950 \& under)'' and ``vehicle stolen'') by different models, specifically concerning the descent and gender of the victims.
}

		\label{Fig2}

	\end{figure*}

BMS (referred to as Structure) achieves accurate detection by utilizing the structural information of behaviors. It ranks first in five indicators and second in the remaining two indicators, as shown in Fig. \ref{Fig2}A. The representation-based models generally lag behind BMS but perform better than observation-based models.
XGBoost, as another ensemble learning method along with GBDT, shows significant performance improvement and almost matches BMS.
This phenomenon suggests that although there is a clear expressive power gap between the theoretical upper bounds of structure-based and observation-based methods, the actual performance achieved by existing models may be far below the ideal performance limit.

In Fig. \ref{Fig2}B, we visualize the trend of changes between the behavior attribute embeddings learned by BMS and the initial semantic embeddings of behavior attributes.
We have clearly tracked three different trends. It's like BMS maintains a vector field in the behavior attribute space, placing each behavior attribute in a position that better aligns with its context.
BMS provides embeddings for each behavior attribute that possesses contextual information based on the known behavior data. Deploying downstream behavior detection tasks on these embeddings significantly improves the detection of challenging behavior data.

In Figs. \ref{Fig2}C and \ref{Fig2}D, we study models' performance under different numbers of behavior attributes. We first witness that, in most cases (when the number is greater than 9), all models show performance gains as the number of behavior attributes increases. This phenomenon can be attributed to the fact that more behavior attributes provide more information, which helps the models accurately differentiate behaviors. This also aligns with the trend of increasing model expressive power. When the number of behavior attributes is small (less than 12), BMS performs lower than representation-based methods.
Facing sparse structural information within a limited number of behavior attributes, representation-based approaches can extract sufficient information for behavior detection even with rough information mining.
 However, as the number of behavior attributes increases, the importance of structural information becomes more pronounced, and representation-based methods gradually lag behind structure-based methods.

Data-driven models may lead to biased outcomes in public domains \cite{mehrabi2021survey}. Specifically, we investigate whether BMS can capture the distribution of criminal behaviors fairly. From Figs. \ref{Fig2}E and \ref{Fig2}F, we find that there is a low correlation between gender and ethnicity with the detected types of criminal behaviors, which is consistent with the original data (figs. \ref{figureS:Crime Data}I and \ref{figureS:Crime Data}J).
Further examining each specific criminal behavior, we note two anomalous phenomena: criminal behavior 1 (``THEFT FROM MOTOR VEHICLE - PETTY (\$950 \& UNDER)'', represented in purple) and criminal behavior 2 (``VEHICLE-STOLEN'', represented in pink) are not proportionally captured by the models. To investigate the cause of these anomalies, we analyze the statistical information regarding different ethnicities and genders of victims in the models' identification of criminal types (Figs. \ref{Fig2}G-\ref{Fig2}J, more information in figs. \ref{figureS:Crime Data}K-\ref{figureS:Crime Data}N).

Regarding criminal behavior 1, the number of correctly identified instances indicates that both the Structure and XGBoost exhibit weak detection capability (the accuracy of only $0.00636$ and $0.04495$). We speculate that this might be due to models misclassifying such behaviors as other criminal types. The detection of these particular criminal behaviors is often considered to have lower reliability.
Regarding criminal behavior 2, the limited number of instances of such behaviors hinders the accurate identification of their criminal types. Similarly, this phenomenon also exists concerning the gender factor of the victims.

\subsection*{Structure does not necessarily ensure gains}

\begin{figure*}[!t]
	\begin{overpic}[height=0.255\textwidth]{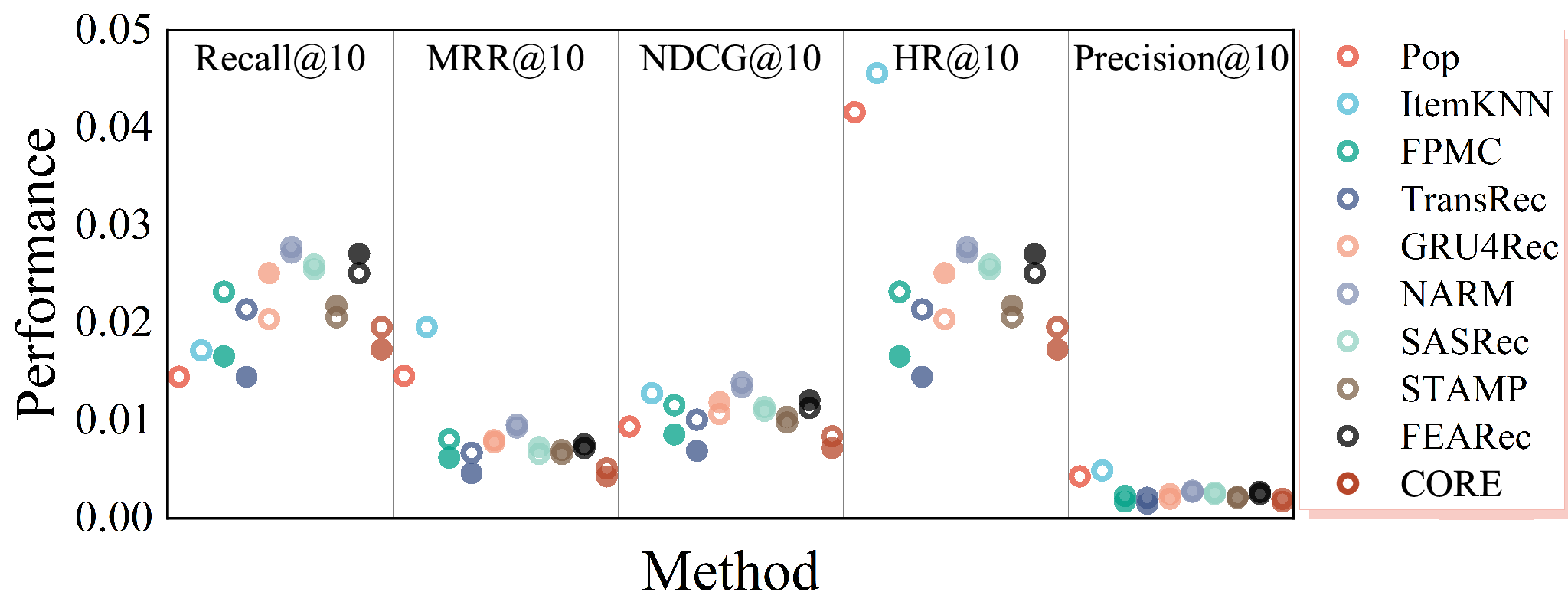}
			\put(-3,34.9){\large\textbf{A}}
		\end{overpic}
		\hspace{2mm}
		\begin{overpic}[height=0.255\textwidth]{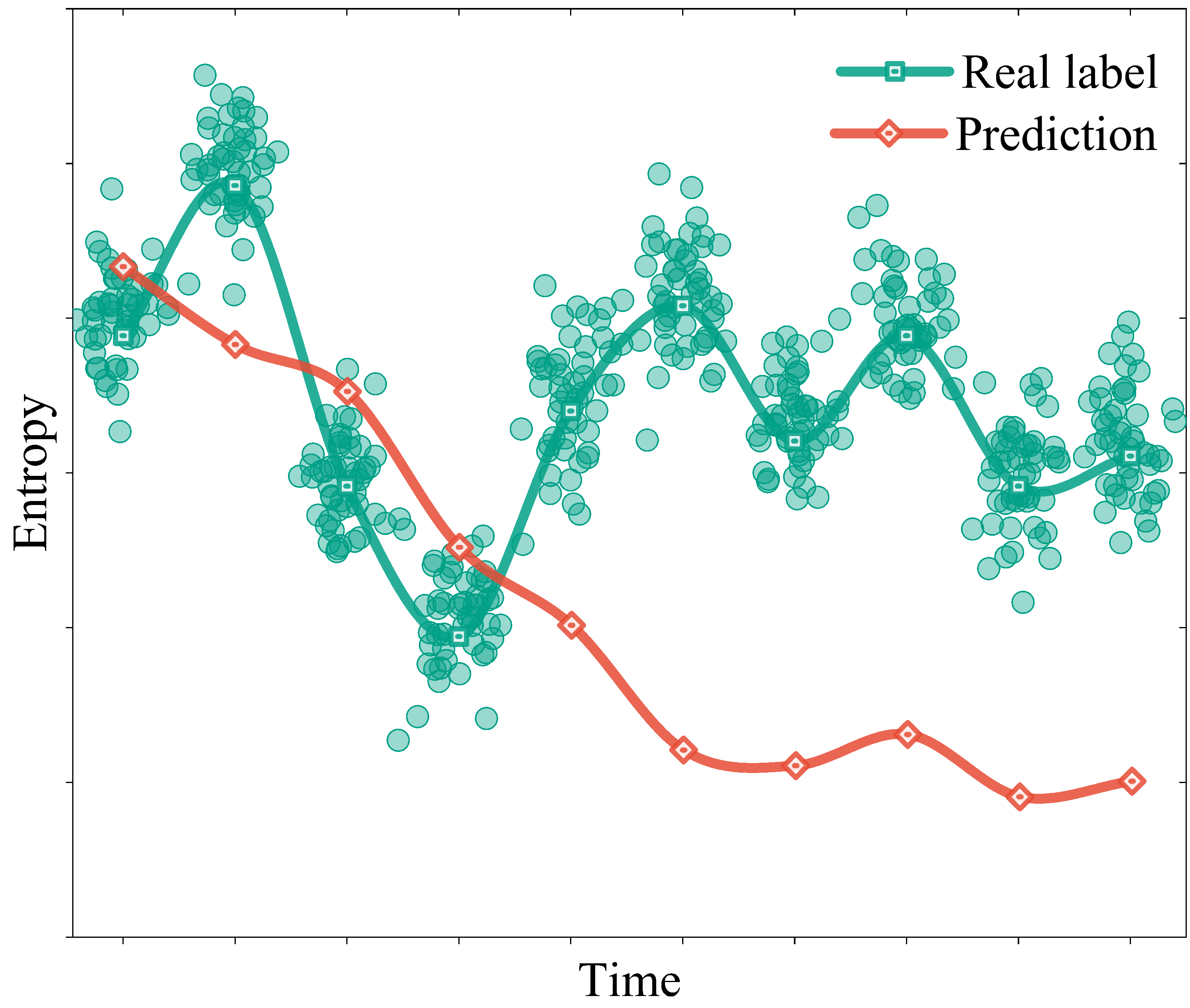}
			\put(-4,76.5){\large\textbf{B}}
		\end{overpic}\\

		\captionsetup[subfigure]{labelformat=empty}
		\caption{\textbf{The thought-provoking performance on behavioral prediction.}
(\textbf{A}) In terms of predicting user interaction behaviors, BMS achieves unstable performance by utilizing behavioral molecular structures that represent multiple recommendation metrics. Hollow points in the figure represent the performance of baseline methods, while solid points represent the performance of baseline methods after leveraging BMS to learn behavioral structures.
(\textbf{B}) To measure the predictability of user behaviors, we recorded the time-independent entropy of randomly selected 200 users in multiple consecutive time intervals and plotted their average value.
}\label{Fig3}

	\end{figure*}

We conduct a study using real user online interaction data to investigate the performance of BMS in behavior prediction. On the knowledge-sharing platform (Zhihu), users are recommended several answers related to questions they are interested in, and they choose to browse the answers they find interesting. By modeling users' interaction behavior based on their historical interactions, we study the prediction of users' potential choices for their next interaction.
The results indicate that BMS exhibits unstable performance in predicting user-interest answers (as shown in Fig. \ref{Fig3}A). For models such as GRU4Rec, STAMP, and FEARec, BMS consistently helps surpass their original performance. Unfortunately, when faced with FPMC, TransRec, and CORE, BMS fails to filter out noise information, leading to the performance decline. The Pop and ItemKNN methods only utilize user-answer interactions for prediction, which prevents BMS from incorporating structural information into these models. Despite using limited information, these two models outperform others in multiple metrics. This phenomenon also confirms that the strength or weakness of model expressive power is not strictly correlated with its performance in specific downstream tasks.

Information entropy, describing the uncertainty of possible behavior occurring in an information source, can be used to capture the predictability of user behavior sequences \cite{doi:10.1126/science.1177170}. We introduce time-independent entropy $S$, $S=-\sum_{j=1}^{N_i}{p_i(j){\log}_2p_i(j)}$, to measure the predictability of user behavior based on their historical interaction records, where $p_i(j)$ represents the historical probability of user $i$ accessing the answer $j$. We randomly select 200 users not included in the training data, analyze their interaction behaviors in consecutive time intervals, and calculate the time-independent entropy. The results are shown in Fig. \ref{Fig3}B. We perceive that with the increase in known behavior, the predicted time-independent entropy shows a decreasing trend, i.e., the uncertainty of user behaviors decreases, making it easier to predict for users with more historical records. Due to the inherent randomness in user behavior sequences, the time-independent entropy of users increases at the second moment instead of showing the expected decreasing trend. But at the third moment, the time-independent entropy of users quickly decreases and even falls below the average value of the predicted results. The predictability of behavior only indicates the probability of correctly predicting a user's future trajectory with an appropriate prediction algorithm.

\subsection*{Structure comes with potential opportunities}

\begin{figure*}[!t]
    \centering
		\begin{overpic}[height=0.185\textwidth]{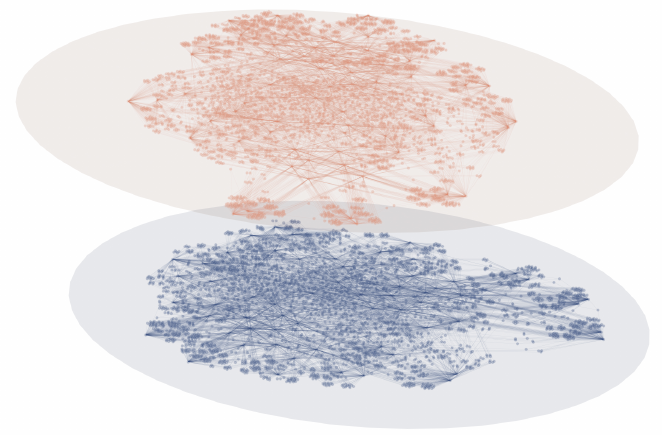}\label{figure:expressive_power}
			\put(-2,64.5){\large\textbf{A}}
		\end{overpic}
		\begin{overpic}[height=0.2\textwidth]{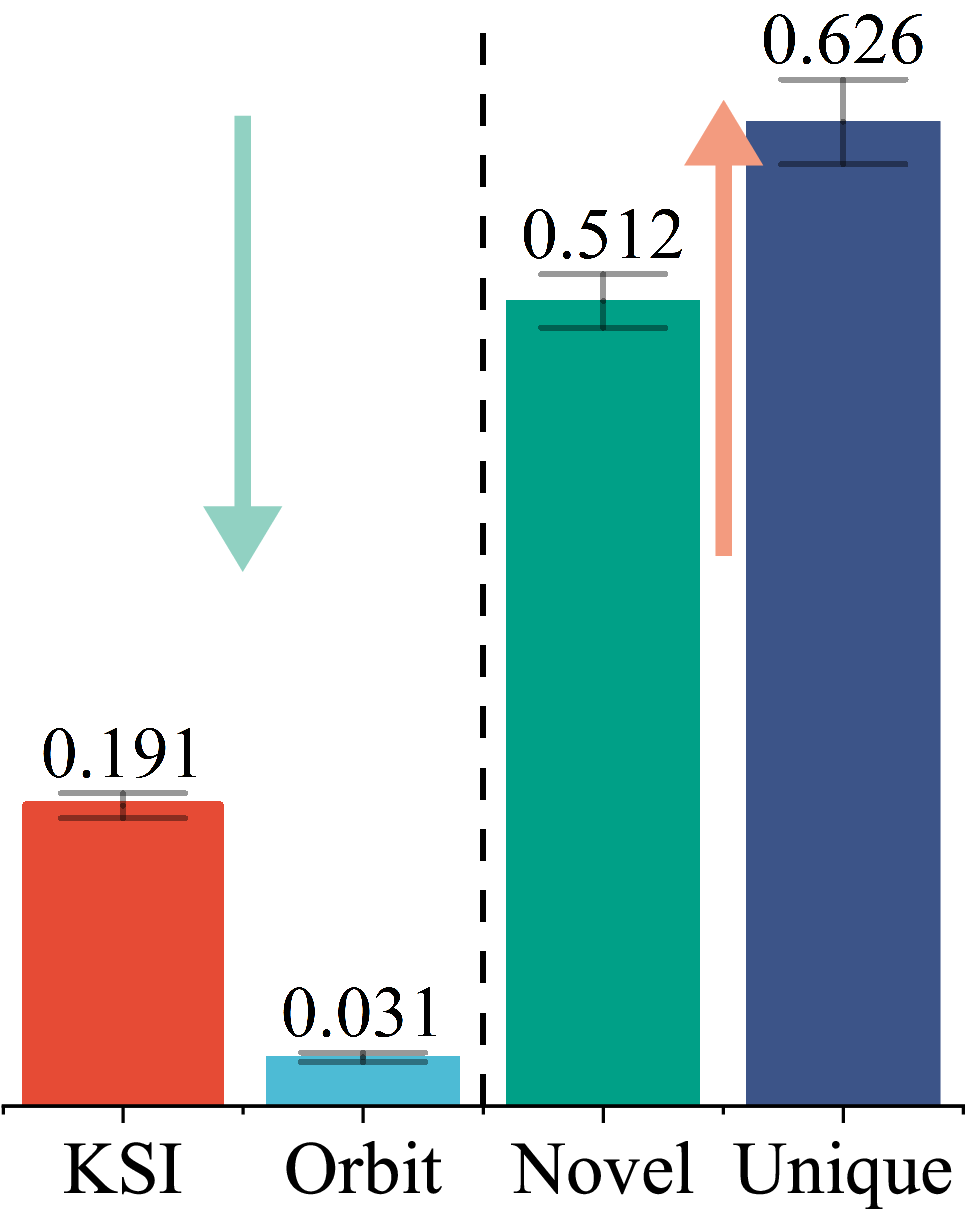}\label{figure:expressive_power}
			\put(-1,90.5){\large\textbf{B}}
		\end{overpic}
		\hspace{1mm}
		\begin{overpic}[height=0.2\textwidth]{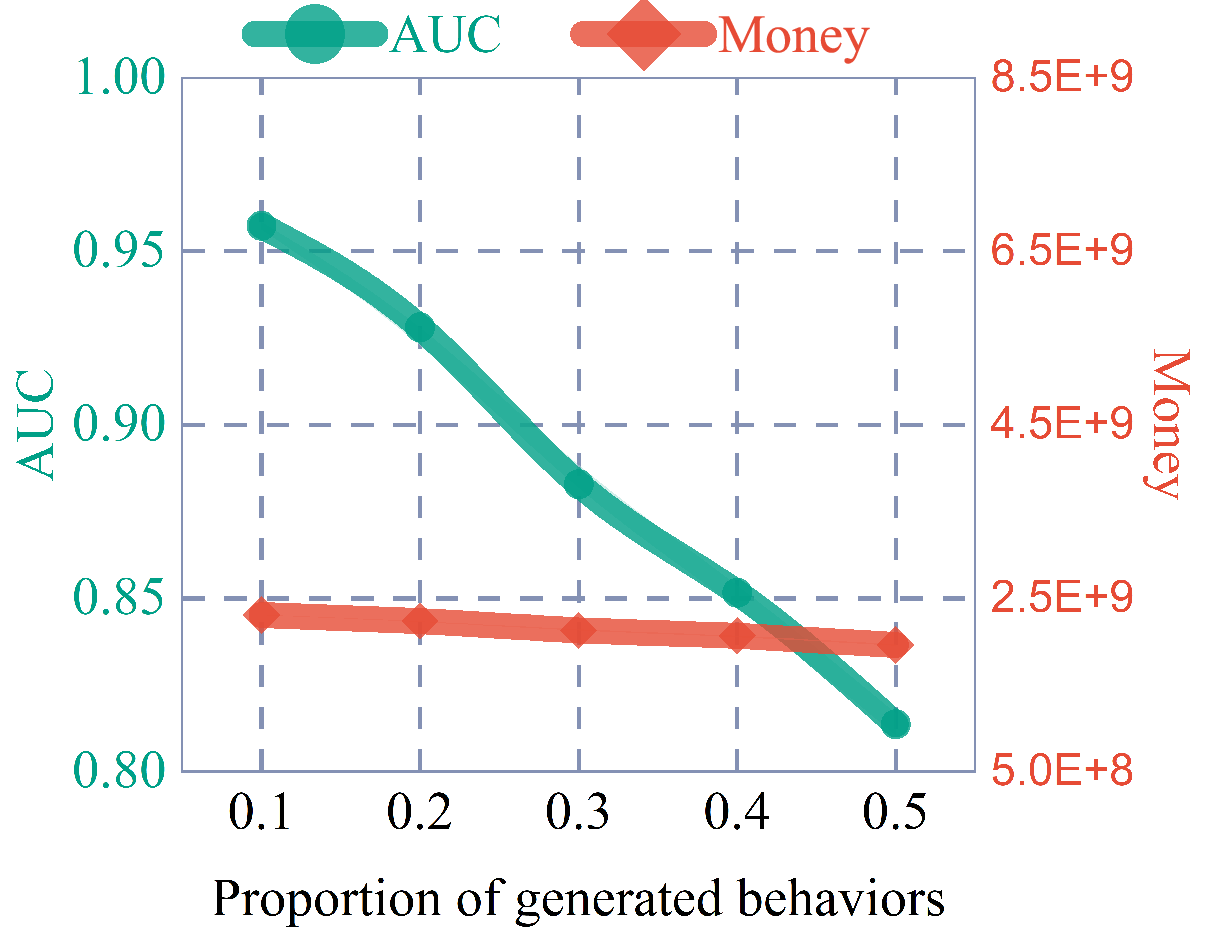}\label{figure:expressive_power}
			\put(-6,70){\large\textbf{C}}
		\end{overpic}
		\begin{overpic}[height=0.2\textwidth]{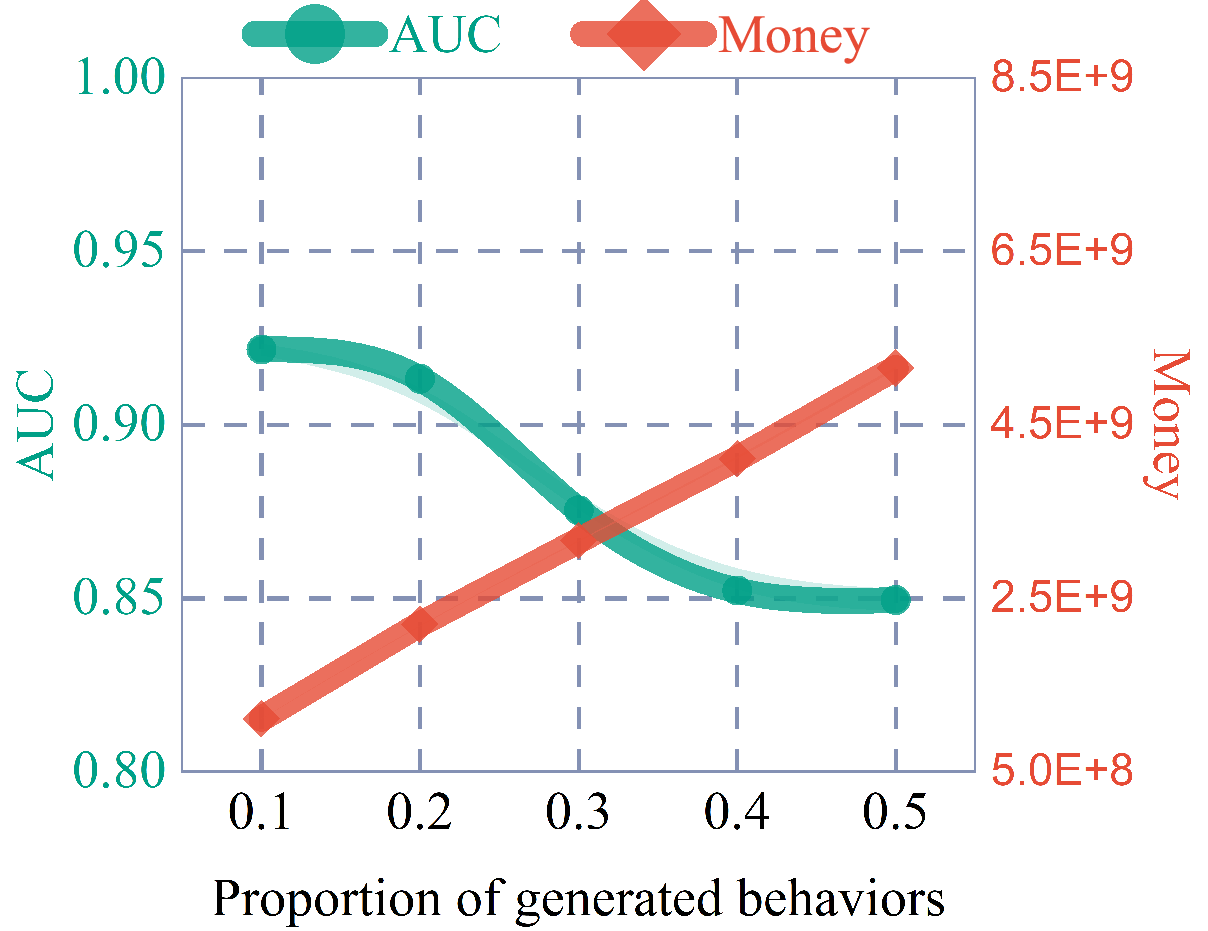}\label{figure:expressive_power}
			\put(-6,70){\large\textbf{D}}
		\end{overpic}

		\captionsetup[subfigure]{labelformat=empty}
		\caption{\textbf{BMS can generate potential reasonable behavioral structures from known behavioral structures and improve the performance of downstream tasks.}
(\textbf{A}) For fraudulent transaction data, based on randomly selecting 2,000 behaviors, BMS generates the behavioral molecular structure (blue) with high similarity compared to the determined structure (orange) in the data, where the generated structure is based only on the limited behaviors in the data (i.e., some behaviors are never applied to train the model).
(\textbf{B}) BMS generates an accurate behavioral molecular structure based on limited behavioral data, which reflects that the generated structure is approximately similar to the original structure from different metrics. Regarding structural evaluation, the left two metrics indicate more similar structures as they become smaller, while a larger value for the right metric indicates discovery of novel structures.
(\textbf{C} and \textbf{D}) For generating invisible behavioral structures, performance decreases moderately with the increasing proportion of invisible behaviors.  In C, we use the generated structures for training, while in D, we use the generated structures for testing.
}

		\label{Fig4}
	\end{figure*}

We deploy BMS to generate behaviors not included in the training set using limited behavioral data for behavior generation and detecting the effectiveness of its generated behavior. We select a fraud transaction dataset that contains over $6$ million records from a bank in South Africa, where fraud behaviors refer to illegal profit from massive transfers from one account to another. Here, part of the behavioral structure will be kept unseen until BMS completes information passing and model training based on known behavioral structures.
We adopt a generative-based method to infer the unseen structures in the behavioral attribute space (Fig. \ref{Fig4}A). In Fig. \ref{Fig4}B, we introduce graph structure-related metrics to compare the known structures with those generated by BMS.
The small KSI and Orbit indicate that BMS has learned approximate structural information. The KSI and Orbit reflect that the generated behavioral structure reserves the expected structure and has a distribution very close to the original structure.
The two other metrics, Novel and Unique, reflect the generalization ability in learning behavioral structure \cite{goyal2020graphgen}. Large values mean that the generated structure contains structures that do not exist in the original data. The results indicate that BMS can effectively capture previously undiscovered structural information.

We further design two groups of experiments to validate the effectiveness of the generated structures by hiding a portion of the raw behavioral data. One is to use the generated structures for the training process, and the other is to use the generated structures for the testing process. Ideally, BMS should generate potential associations among behavioral attributes based on associations with known structures, and these associations can help downstream tasks (Figs. \ref{Fig4}C and \ref{Fig4}D). We observe that the structural information generated from BMS's inference from a part of the known behaviors is less than the structural information contained within the data itself, and the performance of downstream fraud detection decreases with the increasing proportion of hidden behaviors.
In Fig. \ref{Fig4}C, performance continues to decline gradually as more generated structures are used for training,  the decrease in the amount of money involved in fraudulent transactions is minor. It indicates that BMS's ability to learn from small transactions is limited while it can effectively identify key large transactions by capturing crucial structures (similar to functional groups crucial to molecular structures). In Fig. \ref{Fig4}D, as more generated structures are used for testing, we find that the decreasing trend nearly stops when they exceed $40\%$, meaning that most of the generated data are identifiable. Additionally, we note that the prevented fraud losses increase linearly (i.e., downstream models can identify large generated fraudulent behaviors). These experiments show that for high-amount transactions, BMS learns crucial structural compositions in generated structures, while for small-amount transactions, BMS's structural learning is limited due to their scattered, insignificant features.

Overall, we employ behavior molecular structure to provide an expression of behavioral structure, aiming to propose promising signals for the deep utilization of behavioral information. We demonstrate that this solution brings gains in expressive power and utility. It calls for the research community to increase awareness and lay the groundwork for future research on behavioral structure.




\clearpage
\bibliography{scifile}

\bibliographystyle{Science}

\section*{Acknowledgments}
The authors thank Z. Li, T. Hu, Y. Zhao,  Q. Jia, and Y. Wang for discussions and advice.
\textbf{Author contributions:} C.W and C.J designed the research. C.W formulated the idea of behavioral molecular model at atomic-level interrelations. H.Z and Y.H built the initial neural network and performed and analyzed BMS experiments. C.W and H.Z wrote the paper, with input from all authors.
\textbf{Competing interests:} Authors declare that they have no competing interests.
\textbf{Data and materials availability:} All data and code are online available in the supplementary materials.

\end{spacing}

\clearpage
	\graphicspath{{Figures/}{logo/}}

\clearpage
	\graphicspath{{Figures/}{logo/}}

\clearpage
	\graphicspath{{Figures/}{logo/}}

\clearpage
	\graphicspath{{Figures/}{logo/}}
\clearpage
\appendix
\setcounter{table}{0}
\setcounter{figure}{0}
\renewcommand{\thetable}{S\arabic{table}}
\renewcommand{\thefigure}{S\arabic{figure}}

\begin{flushleft}
{\LARGE Supplementary Materials for\\ }

{\textbf{\\ \LARGE On the Expressive Power of Behavior Structure}}

\vspace{0.3in}
{{ \Large Materials and Methods:}}
\end{flushleft}

\section{Behavior description}

In behavioral science, researchers introduce the concept of dimensions to provide a low dimensional metric standard for behavioral expression. Ideally, the dimension of behavior refers to the minimum number of past behavioral attributes needed to make the richest information prediction of future behavior. In this case, the behavioral dimensions are crucial for behavioral expression because the value space of real behaviors is limited by behavioral dimensions, where the differences between different behaviors depend on the values on each dimension and the correlation of values on different dimensions.

In this work, in order to measure the expressive power of different behavioral description methods, we determine the number of different behaviors that the behavior expression can divide in space to reflect its expressive power.
We defined three ways of behavioral expression:

\begin{itemize}
\item[\textbullet] Behavioral observation, which is directly modeled based on observed behavioral attributes.

\item[\textbullet] Behavioral representation, which performs a nonlinear transformation on the original behavioral observations;

\item[\textbullet] Behavioral structure, which characterizes behavior in the behavioral attribute space as a structure composed of attributes and associations.
\end{itemize}

\section{Behavior detection on benchmark dataset 1}
\subsection{Benchmark dataset 1}



To demonstrate the effectiveness of behavioral molecular structure in detection tasks, we utilized a publicly available dataset (Crime Data\footnote{\textbf{Crime Data:} {https://catalog.data.gov/dataset/crime-data-from-2020-to-present}}) consisting of crime incidents in Los Angeles since 2020. Through carefully controlled processing steps, we identified the category of criminal behavior. The code for this section can be found in the ``crime\_detection\footnote{\href{https://github.com/BehavioralComputing/BehaviorStructure/tree/main/crime_detection}{\textbf{Codes link: }https://github.com/BehavioralComputing/BehaviorStructure/tree/main/crime\_detection}}'' folder of our project BehaviorStructure on GitHub. The original data was transcribed from printed crime reports, with the address field containing information up to the nearest hundred blocks to protect privacy. This section presents the raw dataset and describes the preliminary processing steps applied to the raw data.


The raw dataset consists of approximately 771K crime incidents, with each incident recorded in 28 columns. These columns include the date of the incident, the geographical identifier of the police station associated with the case, the type of crime, the gender and ethnicity of the perpetrator, and the location of the crime, among others. For time-related columns such as ``DATE OCC'' and ``Date Rptd'', we split them into three separate fields representing year, month, and day. Furthermore, we extracted the hour and minute from the incident time. Additionally, we introduced a new field called ``Date Difference'', which represents the time difference between the date of the crime and the date of reporting.
The main reason for incorporating the ``Date Difference'' attribute is to explore potential patterns or common characteristics among different types of crimes regarding the time interval between the occurrence of the crime and its reporting, considering the victims' perception and response time. For instance, in the case of severe violent crimes, victims are more likely to report them immediately, resulting in shorter date differences. Conversely, for non-urgent property crimes, victims may require more time to detect and confirm the losses, leading to longer date differences.
To highlight the differences between different behaviors, we removed the fields ``Crm Cd'' (crime code) and ``Crm Cd Desc'' (crime code description), and focused on identifying criminal behavior using the ``Crm Cd'' field as the target for behavior detection. Ultimately, we selected $19$ fields for behavior detection: ``AREA'', ``Rpt Dist No'', ``Part 1-2'', ``Vict Age'', ``Vict Sex'', ``Vict Descent'', ``Premis Cd'', ``Weapon Used Cd'', ``Status'', ``Cross Street'', ``Month\_Rptd'', ``Day\_Rptd'', ``Month\_OCC'', ``Day\_OCC'', ``Date Difference'', ``Hour'', ``Minute'', ``Year\_OCC'', and ``Year\_Rptd''. These fields still include key features such as the timing, location, and victim information within a criminal incident.

To capture the semantic information associated with the values in these fields, we maintained a lookup embedding table. Specifically, we utilized the pre-trained BERT model (``bert-base-uncased'') from Hugging Face library\footnote{\textbf{Hugging Face:} {https://github.com/huggingface}}). Here, we input the string values and leverage the pre-trained deep learning neural network to obtain their corresponding 768-dimensional semantic vectors.

\subsection{Construction of behavioral molecular structure}


We consider the fields in the raw dataset as nodes in the behavior attribute space. For example, there are 12 nodes representing the months, three nodes representing the gender of the victims (with values from the dataset, although our method supports a more diverse gender setting), and several nodes representing the locations and tools used in the incidents. We have defined a meta-rule that specifies the connections between behavior attributes (as shown in Fig. \ref{figure:behavior_graph_construction}A). For example, if an incident occurred on December 1st in Block A, where a Caucasian female was robbed using bare hands, connections would be established between the corresponding nodes for December, 1st, Block A, bare hands, Caucasian, and female. In the entire behavior attribute space, the connections between behavior attributes can be influenced by multiple behaviors. We accumulate these connections, meaning that the edges between behavior attributes that appear more frequently in different behaviors will increase accordingly based on the frequency of occurrence.
\subsection{Implementation of detecting behavioral structures}


In this part, we describe how to learn the behavioral molecular structure obtained from meta-rules, which involves message passing and aggregation between behavior attributes. First, we introduce a linear layer that transforms the 768-dimensional semantic vectors of all behavior attributes into a 128-dimensional vector. Then, we utilize an existing graph neural network method called RGCN (Relational Graph Convolutional Network) to perform message passing and aggregation on the structure composed of all known behaviors.

During each round of message passing in the graph neural network, the RGCN model collects information from neighboring nodes of each behavior attribute node and aggregates this information to update the feature representation of each node. This process consists of three main stages: a) aggregating the features of neighboring nodes, b) updating node features, and c) repeating the message passing.

In stage a), for each node, we aggregate the features of its neighbor nodes by using a summation function to merge the feature vectors of the neighboring nodes. In stage b), we use a gated update function to fuse the aggregated neighbor features with the node's own features and update the node's feature representation. We then repeat stages a) and b) based on the number of layers in the graph neural network.
Once the message passing and aggregation process is completed, we obtain updated node feature representations that incorporate information from the node itself and its neighbors (including first-order and higher-order neighbors).


In the raw data, each behavior corresponds to a subgraph in the behavior attribute space, representing the connectivity patterns among several behavior attributes. We obtain a representation for each behavior by performing average pooling on the node features associated with that behavior. Subsequently, we utilize a three-layer fully connected neural network to map the behavior features to a prediction probability vector. The dimension of this vector depends on the number of behavior categories to be distinguished, where each dimension represents the probability of the behavior vector belonging to a specific behavior category.
During the training process, we use cross-entropy as the loss function and employ the Adam optimization algorithm to minimize the loss function. In this case, the initial learning rate is set to 0.001.

\subsection{Implementation of comparison}


We implemented baseline methods by leveraging scikit-learn and publicly available third-party code from original authors, where applicable. Key parameter configurations were as follows:

KNN: The number of nearest neighbors considered is 5, and the ``brute'' search method is selected.

MLP: The number of layers is 3, with 50 epochs. The activation function is ReLU, and the batch size is 64. The adopted loss function is CrossEntropyLoss from PyTorch, and the optimizer is Adam with an initial learning rate of 0.001.

Tabnet: The number of training epochs set to 10. The patience value used for early stopping, which is 10.

GBDT: \texttt{loss} = log\_loss: The choice of loss function. Here, we use logarithmic loss, also known as logistic loss, to measure the difference between predicted probabilities and true labels. \texttt{learning\_rate} = 0.1: The learning rate controls the contribution of each tree. A smaller learning rate means less influence from each tree, reducing the risk of overfitting. \texttt{n\_estimators} = 5: The number of trees to be constructed, also known as the number of iterations. Here, we set it to 5, meaning 5 trees will be built. \texttt{subsample} = 1: The proportion of the subsample. By default, it is set to 1, indicating the use of the entire training data. \texttt{min\_samples\_split} = 2: The minimum number of samples required to split a node. If the number of samples at a node is less than this value, no further splitting will be performed. \texttt{min\_samples\_leaf} = 1: The minimum number of samples required to be at a leaf node. If the number of samples at a leaf node is less than this value, no further splitting will be performed. \texttt{max\_depth} = None: The maximum depth of the tree. Setting it to None means there is no depth limit for the tree. \texttt{init} = None: The initialization estimator. By default, it is set to None. \texttt{max\_features} = None: The maximum number of features considered when splitting each node. By default, it is set to None, which means all features are considered.
\texttt{verbose} = 0: Controls the verbosity of the output. Setting it to 0 means no process information will be displayed. \texttt{max\_leaf\_nodes} = None: The maximum number of leaf nodes. By default, it is set to None, indicating no limit on the number of leaf nodes. \texttt{warm\_start} = False: Determines whether to continue training from the previous training. Setting it to False means training starts from scratch.

XGBoost: \texttt{objective} = multi:softmax: The choice of objective function. Here, we use the softmax objective function for multiclass problems. It means the model will attempt to output the probability distribution for each class. \texttt{learning\_rate} = 0.1: The learning rate controls the contribution of each tree. A smaller learning rate means less influence from each tree, reducing the risk of overfitting. \texttt{n\_estimators} = 5: The number of trees to be constructed, also known as the number of iterations. Here, we set it to 5, meaning 5 trees will be built. \texttt{max\_depth} = None: The maximum depth of the tree. Setting it to None means there is no depth limit for the tree. \texttt{subsample} = 1: The proportion of the subsample. By default, it is set to 1, indicating the use of the entire training data. If it is less than 1, it means using a portion of the samples. \texttt{colsample\_bytree} = 1: The proportion of features considered when constructing each tree. By default, it is set to 1, meaning all features are considered. \texttt{verbosity} = 0: Controls the verbosity of the output. Setting it to 0 means no process information will be displayed.
\subsection{Behavior analysis}

\begin{figure*}[!t]
	
	\begin{overpic}[height=0.295\textwidth]{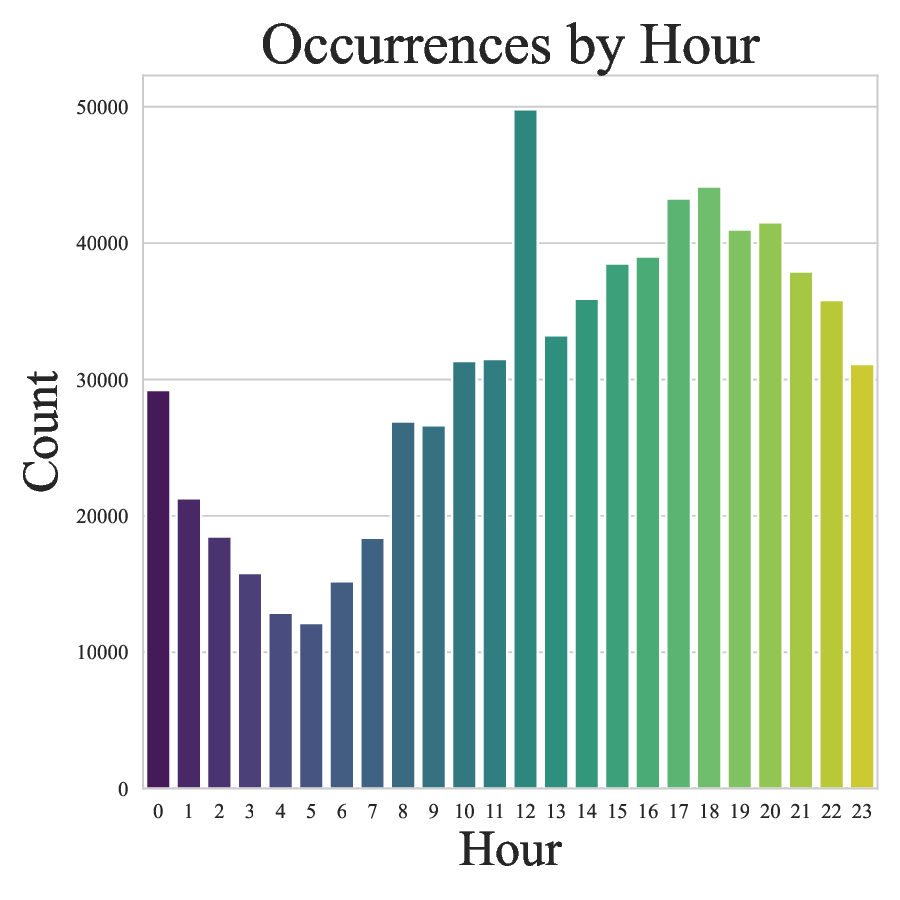}
		\put(-3,80){\large\textbf{A}}
	\end{overpic}
	\hspace{0.08in}
	\begin{overpic}[height=0.295\textwidth]{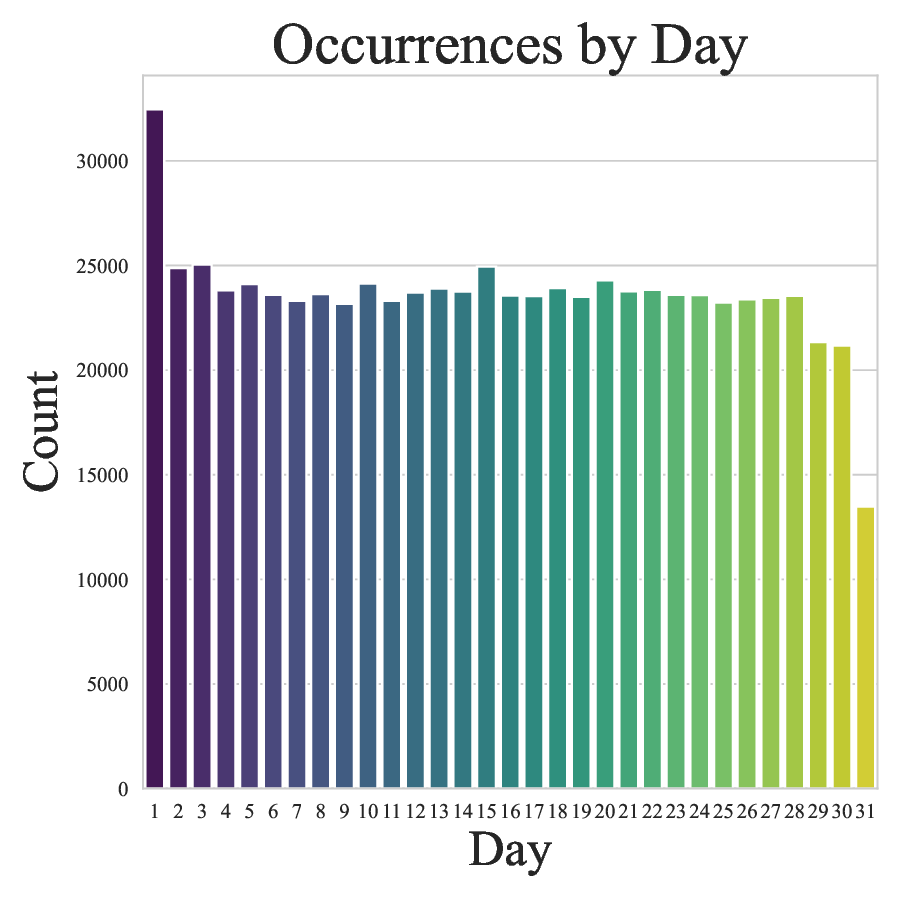}
		\put(-3,80){\large\textbf{B}}
	\end{overpic}
	\hspace{0.08in}
	\begin{overpic}[height=0.295\textwidth]{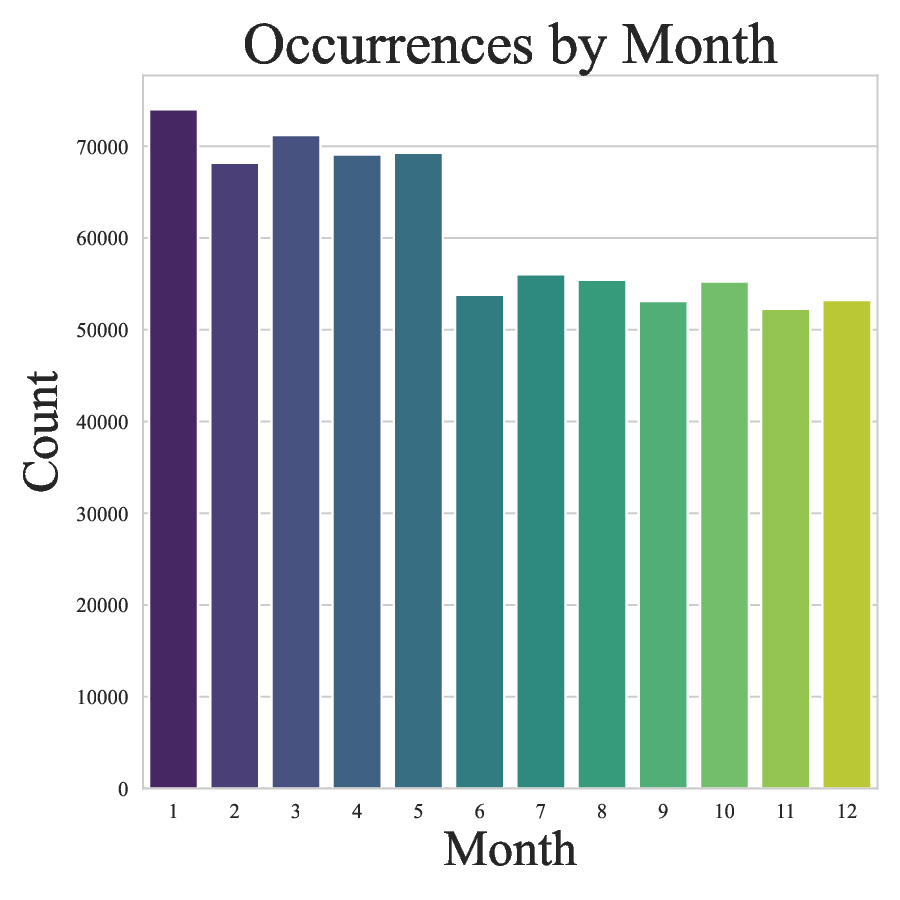}
		\put(-3,80){\large\textbf{C}}
	\end{overpic}

\clearpage	
	\begin{overpic}[height=0.27\textwidth]{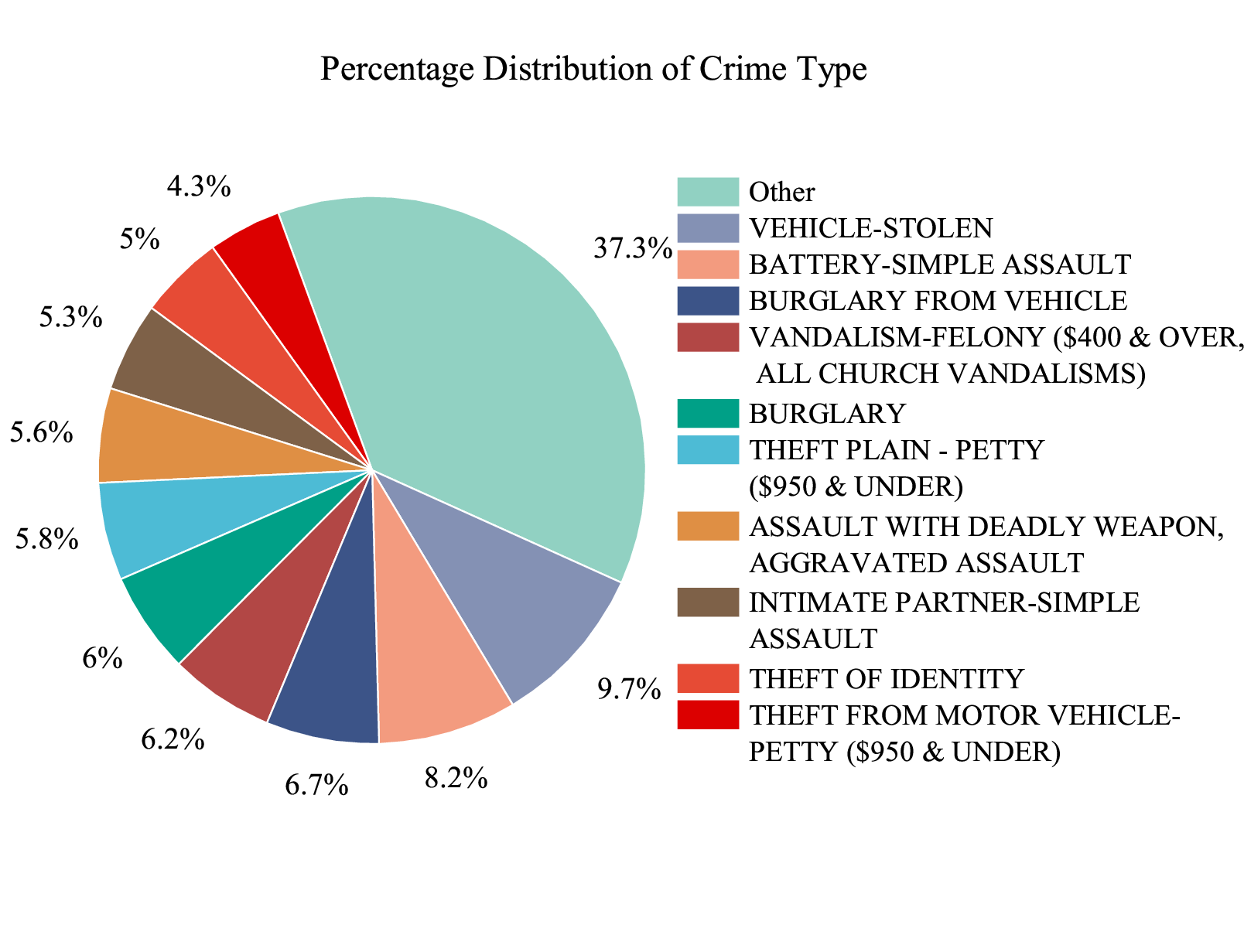}
		\put(-3,69){\large\textbf{D}}
	\end{overpic}
	\hspace{0.045in}
	\begin{overpic}[height=0.27\textwidth]{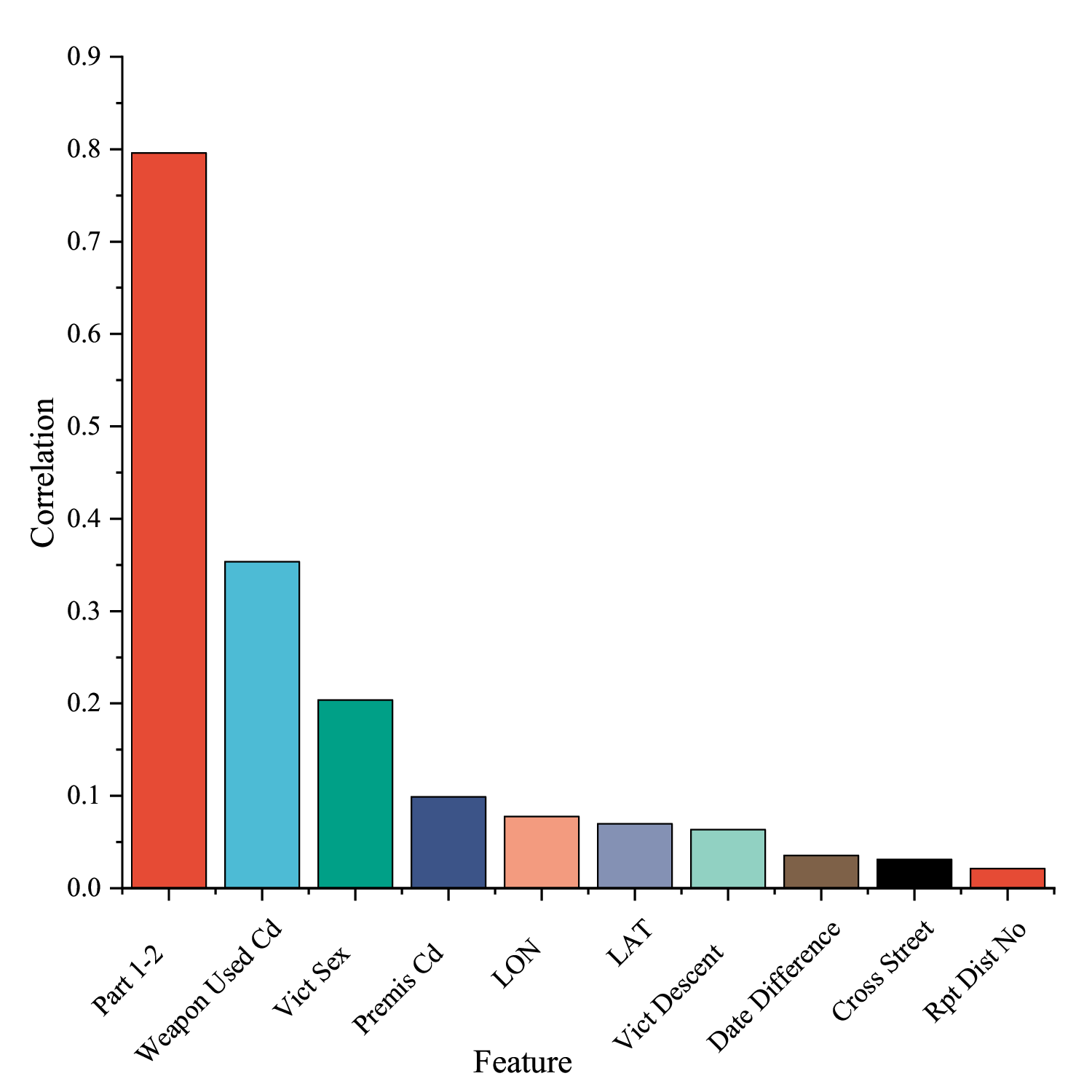}
		\put(-3,90){\large\textbf{E}}
	\end{overpic}
	\hspace{0.045in}
	\begin{overpic}[height=0.27\textwidth]{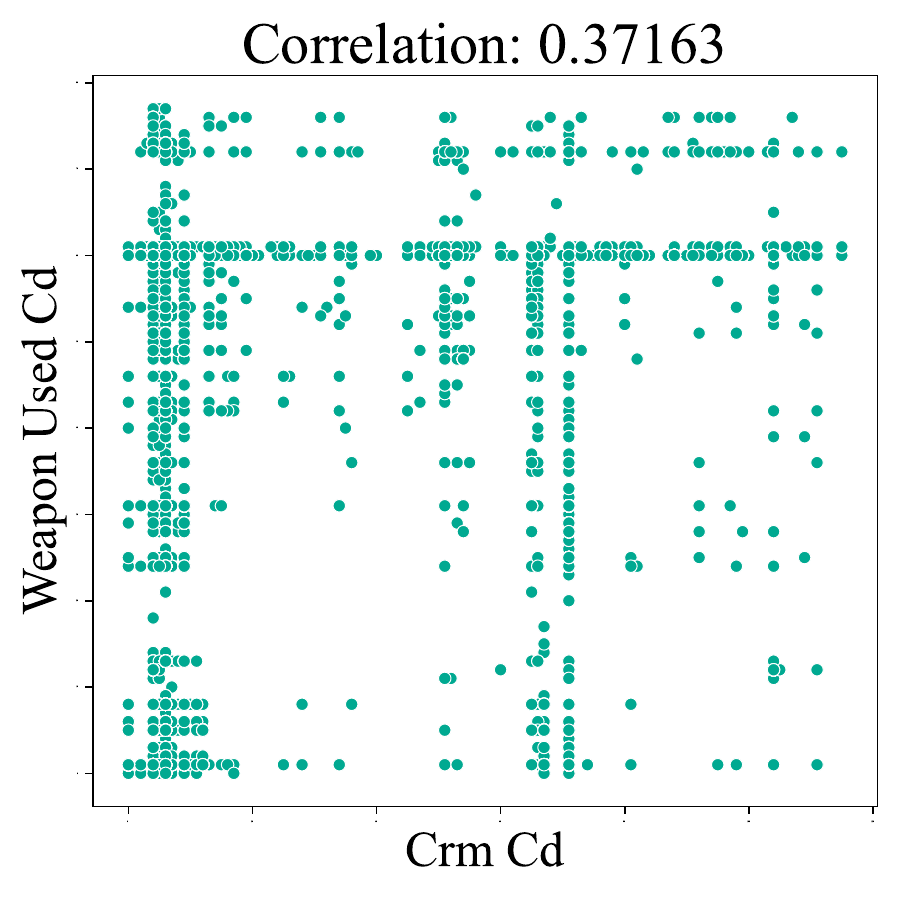}
		\put(-3,90){\large\textbf{F}}
	\end{overpic}

	\begin{overpic}[height=0.23\textwidth]{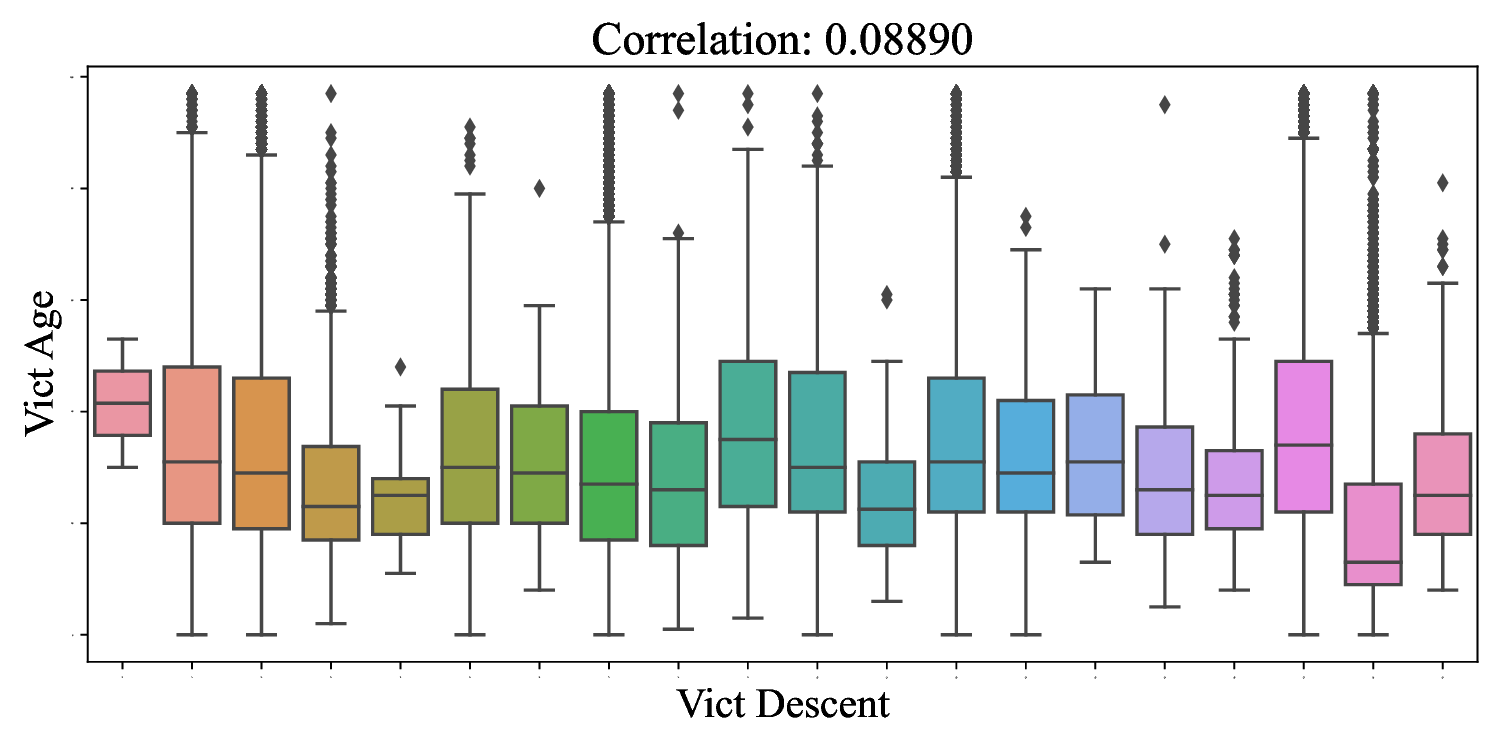}
		\put(-3,40){\large\textbf{H}}
	\end{overpic}
	\begin{overpic}[height=0.23\textwidth]{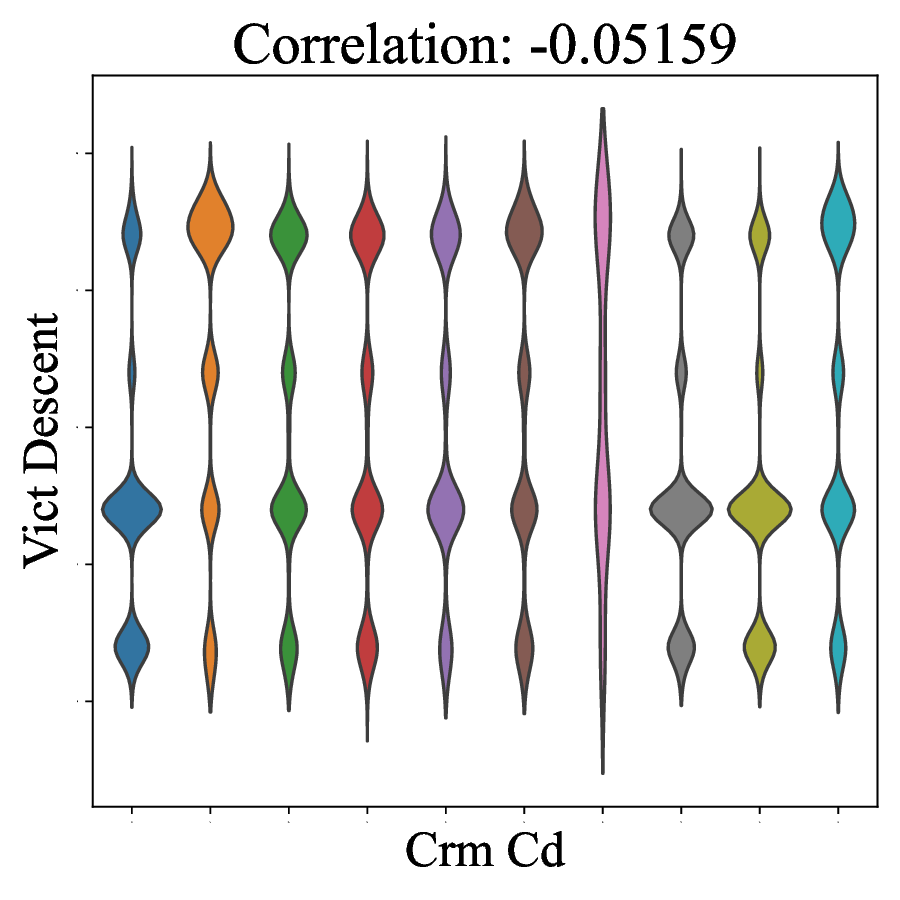}
		\put(-1,80){\large\textbf{I}}
	\end{overpic}
	\begin{overpic}[height=0.23\textwidth]{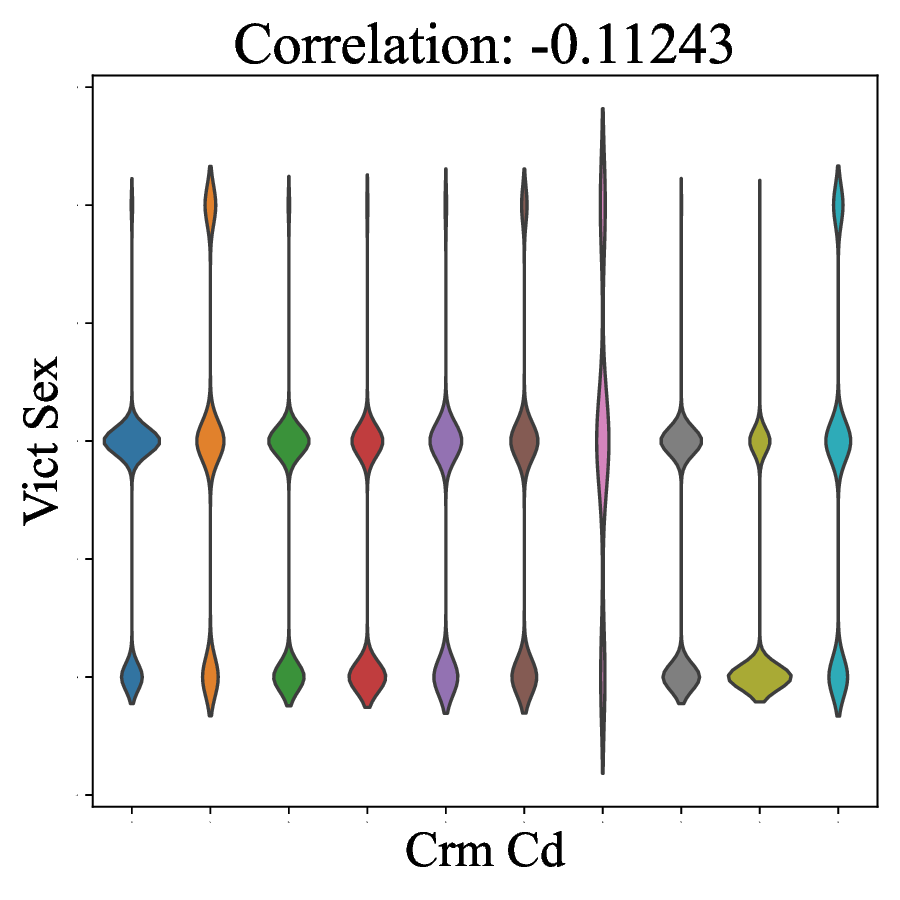}
		\put(-1,80){\large\textbf{J}}
	\end{overpic}

	\begin{overpic}[height=0.175\textwidth]{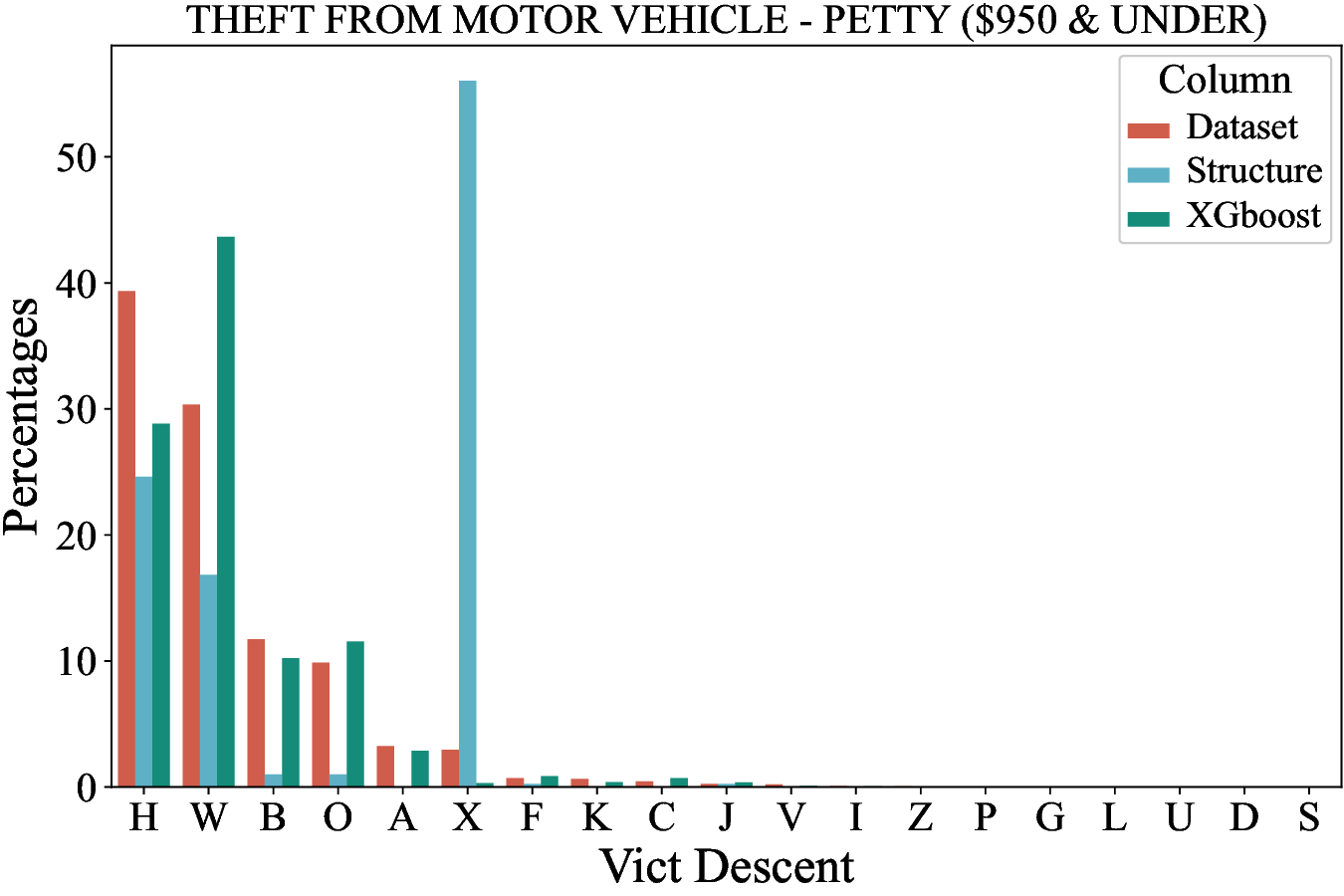}
		\put(-3,55.5){\large\textbf{K}}
	\end{overpic}
	\hspace{1mm}
	\begin{overpic}[height=0.175\textwidth]{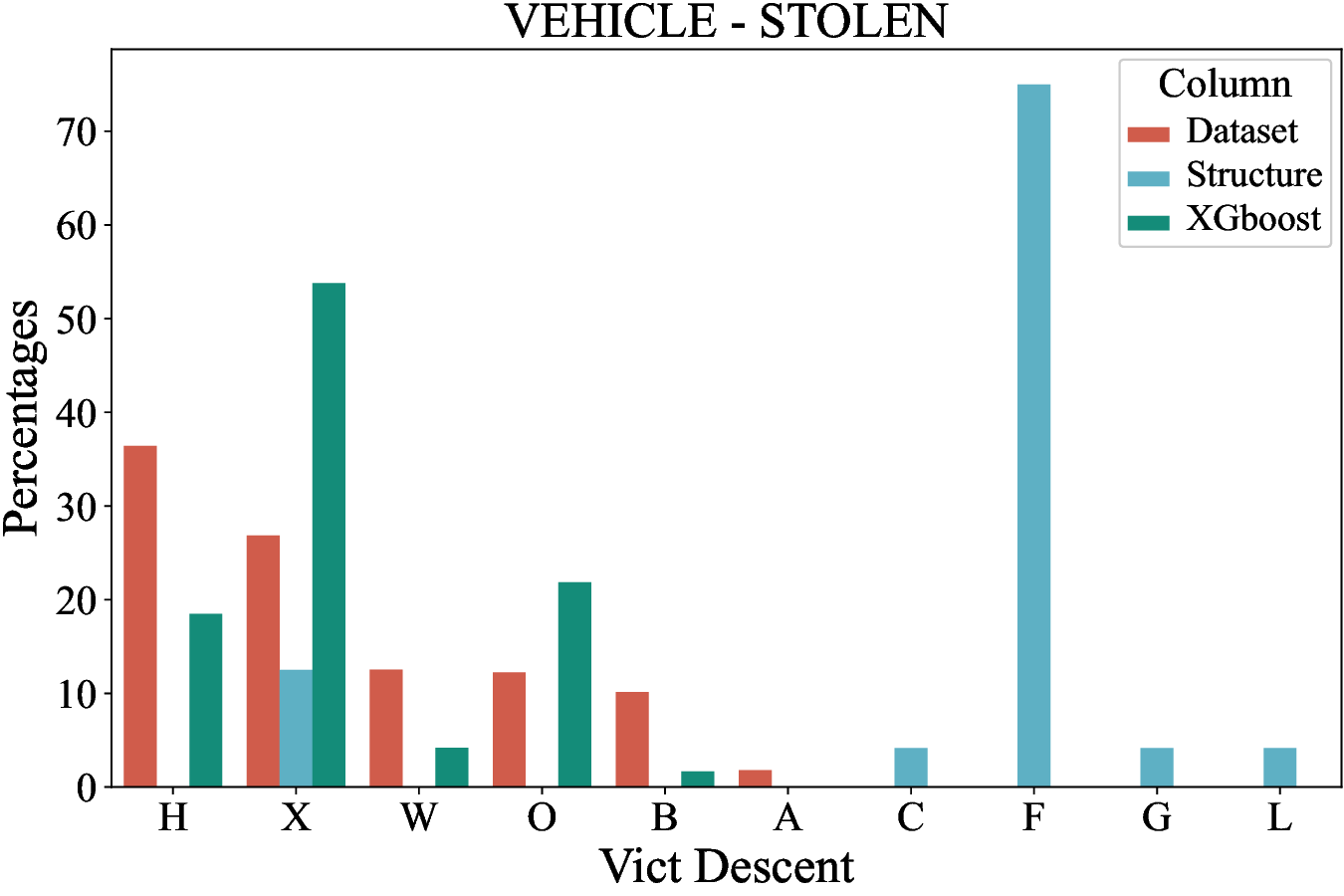}
		\put(-5,55.5){\large\textbf{L}}
	\end{overpic}
	\hspace{1mm}
	\begin{overpic}[height=0.175\textwidth]{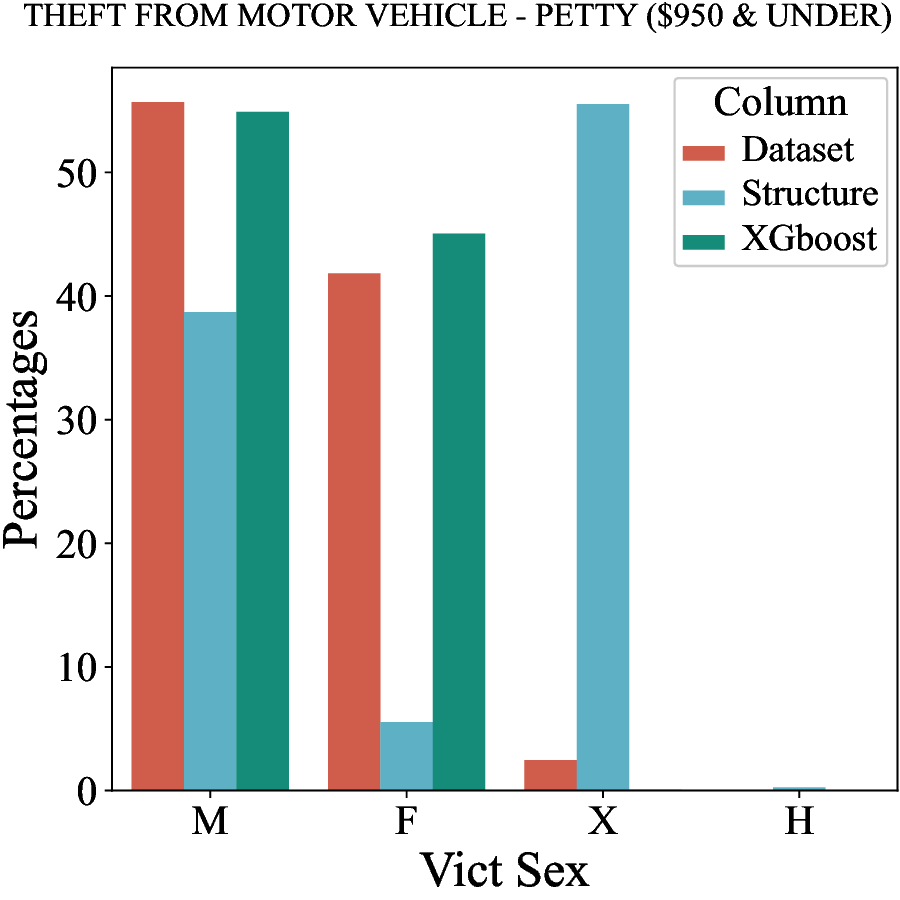}
		\put(-6,83){\large\textbf{M}}
	\end{overpic}
	\hspace{1mm}
	\begin{overpic}[height=0.175\textwidth]{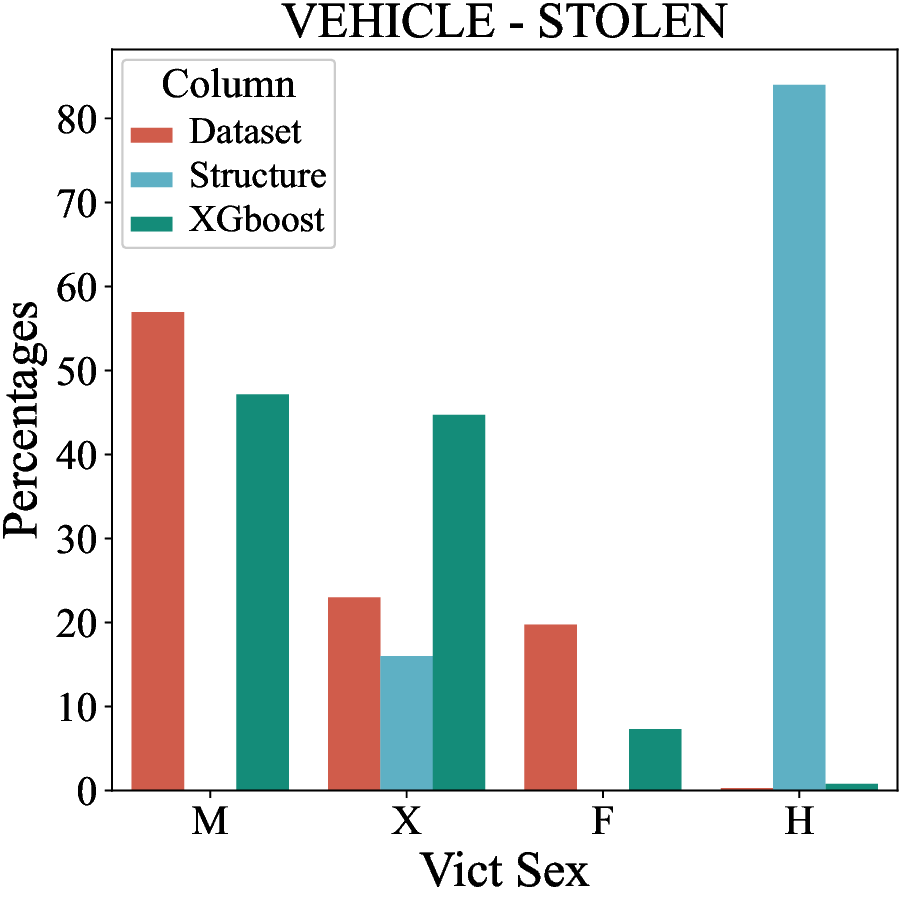}
		\put(-6,83){\large\textbf{N}}
	\end{overpic}

	\captionsetup[subfigure]{labelformat=empty}
	\caption{\textbf{ Detailed analysis of Crime Data. }
	(\textbf{A--C})	display the distribution of crime quantity for each hour in a day, each day in a month, and each month in a year, respectively.
		(\textbf{D}) The proportional distribution of crime categories by quantity in total.
		(\textbf{E}) The top ten features with the highest correlation coefficients to crime categories in the data, with the vertical axis representing the correlation coefficient values.
		(\textbf{F--J}) display the correlation coefficients between crime categories and weapon types, victim descent and victim age, crime categories and victim descent, and crime categories and victim sex, respectively.
		(\textbf{K--N}) display the statistical data regarding the identification of criminal behaviors (``theft from motor vehicle-petty (\$950 under)'' and ``vehicle stolen'') by different models, specifically concerning the descent and gender of the victims.
	}
	
	\label{figureS:Crime Data}
\end{figure*}

We initially conducted an analysis on the temporal distribution of criminal incidents, specifically focusing on the hour, day, and month of occurrence. Through this analysis, we discovered some noteworthy phenomena. In the analysis concerning the hour of occurrence (Fig. \ref{figureS:Crime Data}A), we observed a gradual increase in the number of cases starting from 5 AM, reaching its peak at 6 PM, and then gradually declining. This phenomenon somewhat aligns with human circadian rhythms, as people tend to be more active during the daytime and gradually transition into restful states during the evening. Therefore, the number of criminal incidents also exhibits corresponding trends. It is worth noting that there is a significant increase in the recorded number of incidents at 12 AM. This could be due to a higher concentration of people during this time, such as during lunch breaks, which may potentially trigger more criminal incidents.

In the analysis of the day of occurrence (Fig. \ref{figureS:Crime Data}B), we found a significant increase in criminal incidents on the 1st day of each month, as well as a rise in criminal incidents around the 15th day. In contrast, the number of criminal incidents on other days remains relatively stable. The anomaly in the data for the 31st day (approximately only half the number of criminal incidents) is because not every month has 31 days (only 7 months in a year have 31 days). We tentatively speculate that at the beginning and middle of the month, most employees receive their salaries, leading to increased transaction activities that may include a higher potential for criminal behaviors.

In the monthly analysis (Fig. \ref{figureS:Crime Data}C), the data on criminal incidents across different months are relatively evenly distributed, without any significant variations. However, it is important to note that the dataset covers the period from January 2019 to May 2023, resulting in more data for the months from January to May compared to other months. Overall, the distribution of criminal incidents is more influenced by the hour of occurrence and the day of the month. Considering these temporal patterns will help us better understand the patterns of criminal behavior. Of course, it is also important to conduct more detailed analyses for different types of crime behaviors, as their distributions may reflect different phenomena, which will require further analysis based on downstream tasks.


We conducted a statistical analysis on the quantity of different types of criminal behaviors in Crime Data. The top ten types of criminal behaviors accounted for 62.7\% of the total number of criminal behaviors, as shown in Fig. \ref{figureS:Crime Data}D. The dataset covers 141 different types of criminal behaviors, and the combined quantity of the remaining 131 types only represents 37.3\% of the total. In our crime detection task, we primarily focused on the top ten types of criminal behaviors as our main research subjects. This choice was made because these crime types have higher occurrence frequencies, making them more identifiable and exhibiting lower randomness. Such a selection helps minimize the impact of incidental behaviors on the model, allowing for a better analysis of the structural patterns within criminal behaviors.


We calculated the correlation coefficients between each field in the dataset and the crime type (CRM Cd field) and displayed the top 10 fields with the highest absolute correlation values (Fig. \ref{figureS:Crime Data}E). The reason for the extremely high correlation between the field ``Part 1-2'' and the crime type ``Crm Cd'' is that ``Part 1-2'' is a classification of case information, which is strongly related to the crime type and has only two possible values, 1 or 2. The fields that follow closely are ``Weapon Used Cd'' and ``Vict Sex'', which respectively represent the weapon used during the crime and the gender of the victim. Generally, there is a stronger association between more aggressive weapons and violent crimes, and the female population may be more susceptible to becoming victims of violent crimes. For example, in Fig. \ref{figureS:Crime Data}F, we present a scatter plot that displays the distribution of crime weapons in relation to crime types. We observe a clear correlation between the two variables.


Furthermore, to ensure fairness, we analyzed the relationship between crime types, gender, and ethnicity. In Fig. \ref{figureS:Crime Data}H, we noted that the correlation between the gender and ethnicity of the victims in the included crime behaviors is low (less than 0.09). This suggests that crime behaviors do not strongly target specific ethnicities and genders, and the distribution of victims among different ages and ethnicities is relatively uniform. Moreover, we examined the relationship between crime types and gender (Fig. \ref{figureS:Crime Data}I), as well as crime types and ethnicity (Fig. \ref{figureS:Crime Data}J). We noted that there is almost no correlation between crime behaviors and victim ethnicity (only about 0.05), while there is a slightly higher correlation between victim gender and crime behaviors. This is partly due to certain crime behaviors, such as sexual assaults, occurring more frequently against specific genders.

\section{Behavior prediction on benchmark dataset 2}
\subsection{Benchmark dataset 2}




The ZhihuRec Data\footnote{\textbf{ZhihuRec Data:} {https://github.com/THUIR/ZhihuRec-Dataset}} is a real-world personalized recommendation interaction dataset used exclusively for scientific research. It was jointly constructed by the Information Retrieval group of Tsinghua University (THUIR) and Zhihu company (a knowledge-sharing platform). In our work, we focused on studying the performance of behavior substructure in behavior prediction based on the ZhihuRec dataset.
We have published the code for this section in the ``behavior\_prediction\footnote{\href{https://github.com/BehavioralComputing/BehaviorStructure/tree/main/behavior_prediction}{\textbf{Codes link: } https://github.com/BehavioralComputing/BehaviorStructure/tree/main/behavior\_prediction}}'' folder of our project BehaviorStructure on GitHub.

The dataset consists of approximately 100 million interaction logs, capturing various user, answers, questions, authors, and topics within the Zhihu question and answer community. In this paper, we first performed data filtering and aggregation on the original dataset, and then deployed the behavioral molecular structure model based on the newly constructed fields.

\subsection{Construction of behavioral molecular structure}

We primarily utilize the ``info\_user.csv'', ``info\_answer.csv'', and ``inter\_impression.csv'' files from the dataset, representing user information, answer information, and user-answer interaction behavior data, respectively. The specific extracted fields are shown in Table \ref{Table:dataset zhihu}.
Based on the ``inter\_impression.csv" file, we established the interaction relationship between user IDs and answer IDs to establish the connection between users and answers. Furthermore, in the ``info\_answer.csv" file, we extracted the one-to-one correspondence between answer IDs and answer contexts. Additionally, when an answer appears on the user interface, a timestamp is generated to record this event (corresponding to ``impression timestamp'' in the inter\_impression.csv' file). Furthermore, when a user accesses an answer, the timestamp of the user's click is recorded in the ``inter\_impression.csv'' file (denoted as ``click timestamp"). With this, we can establish the association between a user accessing an answer at a specific moment.
To avoid having too many fields, we binned the ``click timestamp'' and ``impression timestamp'' fields by hour. Similarly, for the ``register timestamp'' field in the ``info\_user.csv'' file, we binned it by month.
Both answer topics and user topics have the same value range, consisting of topic IDs. Therefore, we consider them as a single field called ``topic", and we can establish the association between a user accessing an answer under a specific topic.
Taking into account the relevant information for users and answers, we establish associations with the user ID and answer context, respectively.
In summary, when a user initiates a click on an answer, its corresponding substructure can be modeled as a graph structure (as shown in Fig. \ref{figure:behavior_graph_construction}B).

\subsection{Implementation of predicting behavioral structures}


In the behavior prediction, we employ graph neural networks as the learning module for behavior molecular structure. It consists of two main components: feature embedding and message passing. The purpose of feature embedding is to map the original data into a vector space. The model initializes several feature embedding layers for each feature node, the number of which depends on the number of types for each node. The input raw features are passed through the feature embedding layers, resulting in low-dimensional vectors in the representation space. Specifically, we used a two-layer RGCN layer with an activation function of Relu, and a linear layer to obtain the final output embeddings. Based on the distribution of user clicks, we classify users into $4$ categories based on clicks of $0$-$19$, $20$-$49$, $50$-$99$, $100$ and above, as labels for node classification when learning output embeddings. In this work, we set the embedding dimension to be 64.

\begin{figure*}[!t]
	\centering
	\begin{overpic}[width=0.50\textwidth]{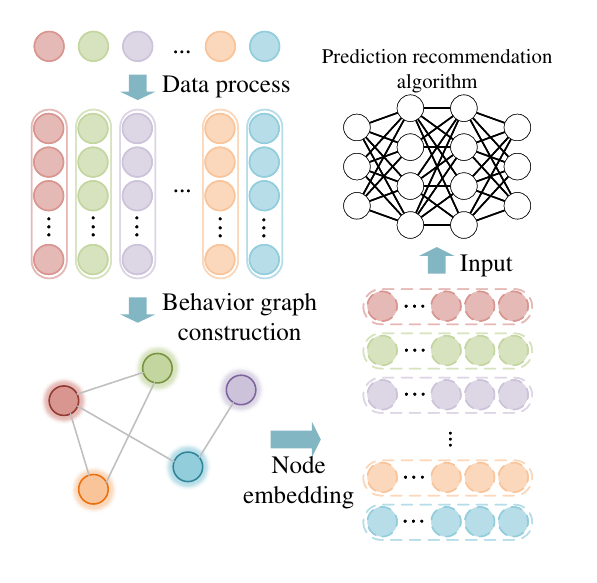}
		\put(-1,80){\large\textbf{ }}
	\end{overpic}

	\captionsetup[subfigure]{labelformat=empty}
	\caption{\textbf{The behavioral structure prediction framework.}\label{figure:behaviorprediction_workflow}
	}
	
\end{figure*}

After obtaining the embeddings of the nodes, we extract 64-dimensional vectors for each user ID and answer ID, which serve as features for subsequent tasks. For the subsequent task, we utilize the RecBole\footnote{\textbf{RecBole:} https://recbole.io} platform and employ its sequential recommendation method as our prediction approach.
Specifically, leveraging the interaction data and features of users and answers, we recommend answers to users using the sequential recommendation method. This entails predicting whether a user will click on a specific answer. If the user indeed clicks on the answer, it is considered a correct prediction.
By utilizing the RecBole platform, we leverage its sequential recommendation method to make predictions based on the user-answer interactions and features. This allows us to effectively predict the behavior of users, specifically whether they will click on a given answer, and assess the performance of our model.
The overall workflow is illustrated in Fig. \ref{figure:behaviorprediction_workflow}.

%
%
%
%

%

\subsection{Implementation of comparison}

In this experiment, we select four metrics, Recall@10, MRR@10, NDCG@10, Hit@10, and Precision@10, to evaluate the performance of the models.

Recall@10: Recall is one of the metrics used to measure the coverage of a recommendation system. Recall@10 measures the proportion of items that appear in the top 10 positions of the recommendation list, among the items that the user has actually interacted with. The value of Recall@10 ranges from 0 to 1, with higher values indicating a stronger ability of the recommendation system to cover the user's actual interests.

MRR@10 (Mean Reciprocal Rank at 10): MRR is a metric used to evaluate the ranking quality of a recommendation system. MRR@10 calculates the average reciprocal rank of the first item that the user has actually interacted with in the recommendation list. Reciprocal rank refers to the reciprocal of the rank of an item, for example, the reciprocal rank of the item at the first position is 1, the reciprocal rank of the item at the second position is 1/2, and so on. The value of MRR@10 ranges from 0 to 1, with higher values indicating better performance of the recommendation system in ranking items that the user is interested in.

Normalized Discounted Cumulative Gain, NDCG, is similar to HR but additionally considers the position of the user's accessed item in the recommended list.
The calculation formula for NDCG is as follows:

$$\mathrm{NDCG} = \frac{1}{N}\sum\nolimits_{i=1}^{N}\frac{1}{\log_2{{(P}_i+1)}},$$
where $N$ represents the total number of users, and $P_i$ represents the position of the accessed item of user $i$ in the recommended list. If the item does not exist in the list, $P_i$ approaches infinity.

Hit Ratio@K, also known as Top-K Hit Ratio, is a commonly used evaluation metric in recommendation systems. It measures the percentage of the top 10 or top 20 recommended items in the recall list that match the user's actual accessed items, reflecting the model's performance in recommending the top-K items. The calculation formula for HR is as follows:

$$\mathrm{HR} = \frac{1}{N}\sum\nolimits_{i=1}^{N}{hits(i)},$$
where $N$ represents the total number of users, and $hits(i)$ indicates whether the user's accessed item is in the recommended list (1 for yes, 0 for no).

Precision@10: Precision is another metric used to measure the accuracy of a recommendation system. Precision@10 calculates the proportion of items in the top 10 positions of the recommendation list that the user has actually interacted with. The value of Precision@10 ranges from 0 to 1, with higher values indicating more accurate recommendations in the top 10 positions of the recommendation list.

We select many well-established and high-performing algorithms for comparison, including Pop, ItemKNN, BPR, LightGCN, and ENMF.
These algorithms represent both traditional and session-based recommendation approaches. The purpose is to demonstrate the effectiveness of the proposed algorithm by comparing its performance against these established algorithms.

\section{Behavior generation on benchmark dataset 3}

\subsection{Benchmark dataset 3}
For the sake of validating BMS's generative capabilities, we selected the Fraudulent Transaction Data\footnote{\textbf{Fraudulent Transaction Data:} {https://www.kaggle.com/datasets/chitwanmanchanda/fraudulent-transactions-data}}, which is available publicly on the Kaggle platform. The dataset contains 6,362,620 behavioral transactions consisting of 10 fields (Table \ref{Table:dataset fraudulent}), which involve transfers between bank accounts. We performed the following preprocessing steps on the original data. Since the field step is naturally a discrete integer representing the time period encoding of when the transaction occurred, no further processing is needed. For string-type fields such as type, nameOrig, and nameDest, inspired by natural language processing, we converted the original text data into tagged format (i.e., labeled processing). For amount-type fields such as amount, oldBalanceOrg, newbalanceOrig, oldbalanceDest, and newbalanceDest, we first labeled them and then bucketed them according to their maximum counting unit in decimals.
We have published the code for this section in the ``graph\_Gen\footnote{\href{https://github.com/BehavioralComputing/BehaviorStructure/tree/main/graph_Gen}{\textbf{Codes link: } https://github.com/BehavioralComputing/BehaviorStructure/tree/main/graph\_Gen}}'' folder of our project BehaviorStructure on GitHub.

\subsection{Construction of behavioral molecular structure}


In this study, we use graph structures to represent behavioral molecular structures. We map each behavior to a subgraph structure consisting of a set of nodes and edges, where different fields in the original behavioral dataset are considered as different types of nodes. Therefore, each behavior can generate a corresponding heterogeneous graph. Based on our understanding of anti-fraud business processes, we define meta rules by selecting corresponding fields to represent nodes and selecting specific node pairs to represent edges, thus generating behavioral molecular structures containing 11 different types of nodes and 11 different types of edges (as shown in Fig. \ref{figure:behavior_graph_construction}C).

\subsection{Implementation of generating behavioral structures}

\begin{figure*}[!t]
	\centering
	\begin{overpic}[width=0.90\textwidth]{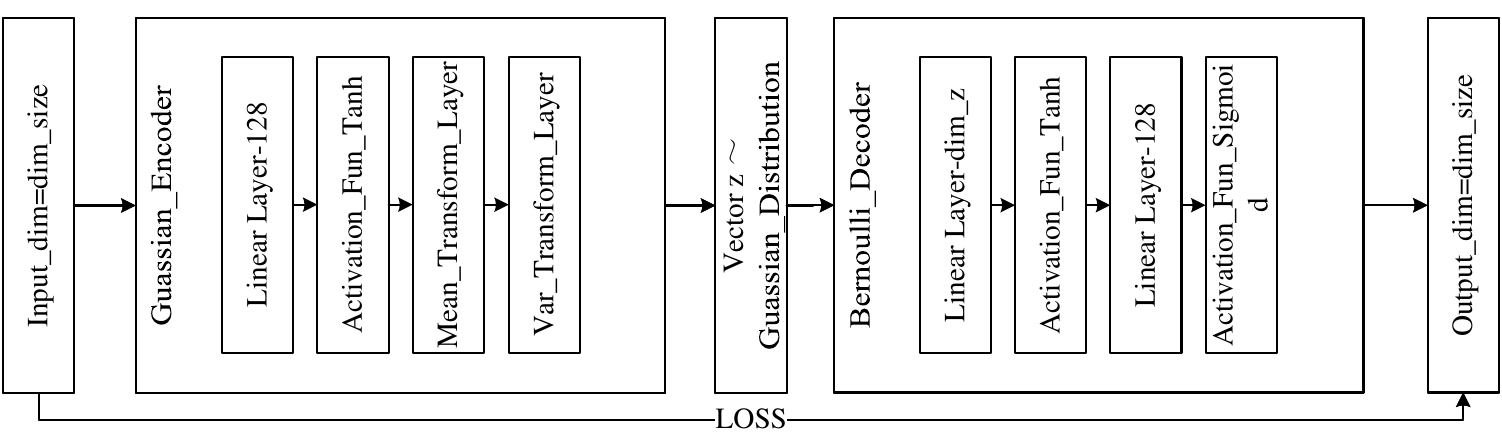}\label{figure:behaviorgeneration_framework}
		\put(-1,80){\large\textbf{ }}
	\end{overpic}

	\captionsetup[subfigure]{labelformat=empty}
	\caption{\textbf{The behavioral structure generation framework based on GraphVAE.}\label{figureS:graphsageframework}
	}
	
\end{figure*}
\begin{figure*}[!t]
	\centering
	\begin{overpic}[height=0.285\textwidth]{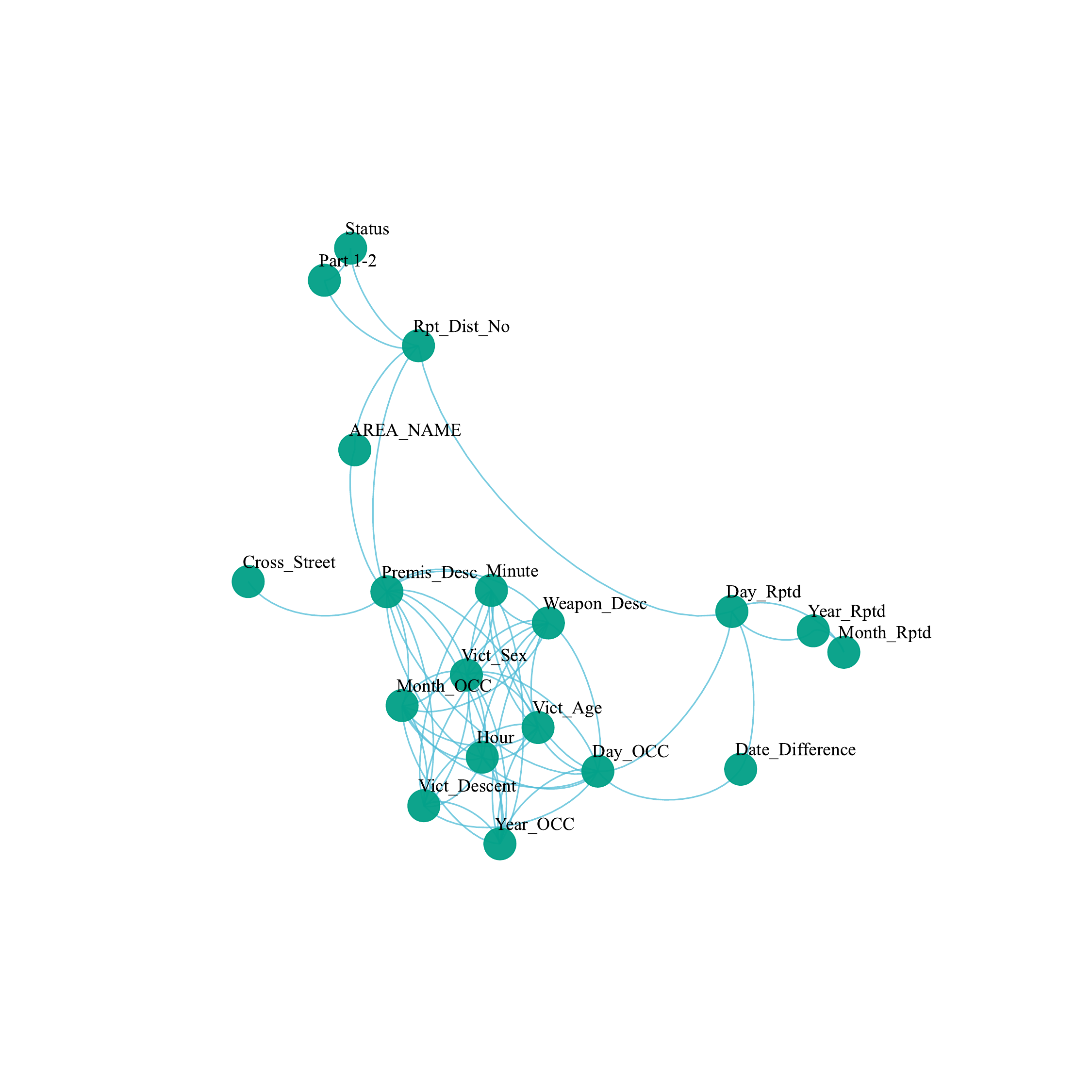}
		\put(-1,85){\large\textbf{A}}
	\end{overpic}
	\begin{overpic}[height=0.285\textwidth]{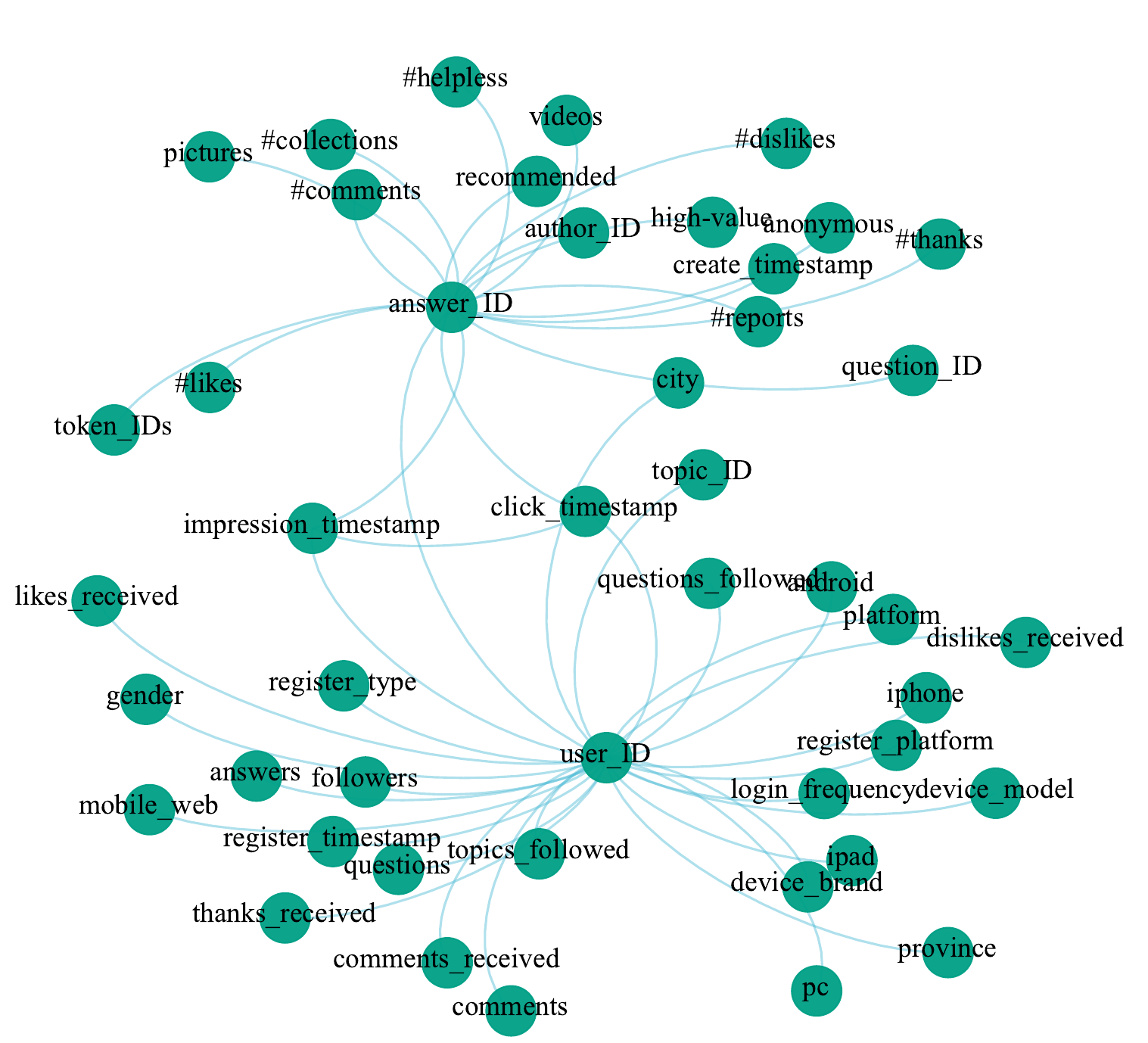}
		\put(-1,77){\large\textbf{B}}
	\end{overpic}
	\begin{overpic}[height=0.285\textwidth]{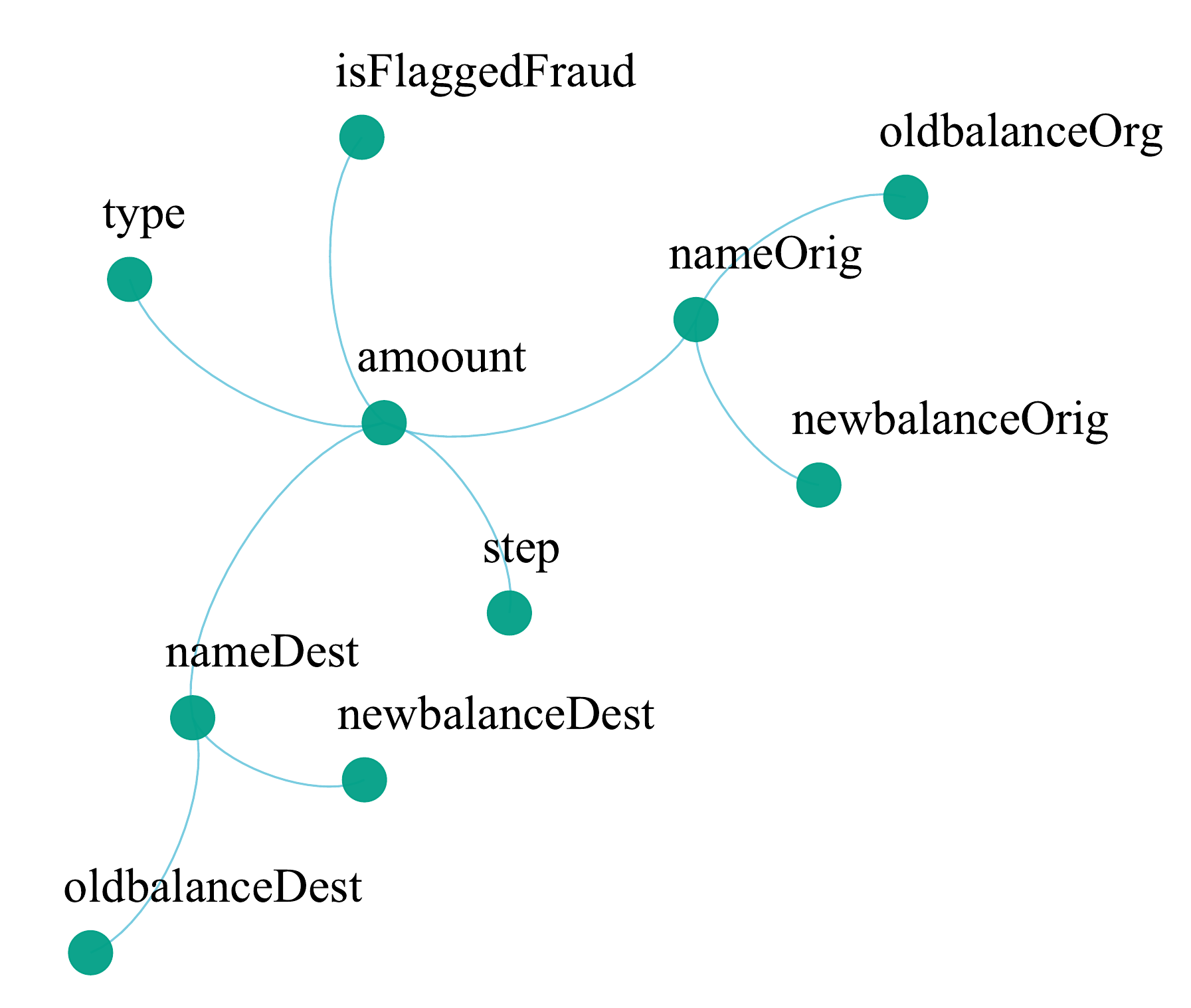}
		\put(-1,72){\large\textbf{C}}
	\end{overpic}

	\captionsetup[subfigure]{labelformat=empty}
	\caption{\textbf{Construction of behavioral molecular structure on one behavior data.}
		(\textbf{A}) Behavior detection in Crime Data.
		(\textbf{B}) Behavior prediction in ZhihuRec Data.
		(\textbf{C}) Behavior generation in Fraudulent Transaction Data.
		The nodes represent the values of behavior attributes, and the edges indicate the relationships between two behavior attributes based on our custom-defined meta-rules.
	}\label{figure:behavior_graph_construction}
\end{figure*}

The steps for behavioral structures generation that we utilize an existing method called GraphVAE. The framework of structures generation is illustrated in Fig. \ref{figureS:graphsageframework}.

%
%
%
%

Assume we have a graph $G=(A,E,F)$, with adjacency matrix $A$, edge attributes $E$ and node attributes $F$. We want to learn an encoder and decoder to map the graph $G$ and embedding $\vec{z}\in \mathbb{R}^c$. In the GraphVAE probabilistic setting, the encoder is a variational posterior distribution $q_\Phi(\vec{z}|G)$, the decoder is the generative distribution $p_\theta(G|\vec{z})$, and $\vec{z}$ is assumed to have a standard Gaussian prior. We want to minimize the negative log likelihood $-\log{p_\theta(G)}$, according to GraphVAE derivations, we have the following loss function:
\begin{equation}
\mathcal{L}(\phi, \theta, G)=\mathbb{E}_{q_{\phi}(\vec{z} \mid G)}\left[-\log p_{\theta}(G \mid \vec{z})\right]+K L\left[q_{\phi}(\vec{z} \mid G) \| p(\vec{z})\right].
\end{equation}
The first term of the loss function is called the reconstruction loss. This term forces the generated graph to be highly similar to the input graph $G$. The second term is computed based on the KL divergence, which allows the encoder to approximate the prior distribution. Typically, the dimension of $\vec{z}$  is relatively small, which encourages the autoencoder to learn a high-level compression of the input rather than simply copying any given input. In the following, we will introduce our graph decoder and the appropriate reconstruction loss.

Given an input graph $G$ with $n \leq k$ nodes and the corresponding reconstructed graph $\widetilde{G}$, calculating the loss function requires evaluating the likelihood function, which measures how likely graph $\widetilde{G}$ is generated from graph $G$:

\begin{equation}
p_{\theta}(G \mid \vec{z})=p(G|\widetilde{G}) = {\frac{p(\widetilde{G}|G) p(G)}{p(\widetilde{G})}}.
\end{equation}

Since there is no specific node order in graphs and the adjacency matrix of a graph is invariant to node permutation, comparing two graphs for isomorphism is challenging. However, we can obtain a binary assignment matrix $X\in\{0,1\}^{k\times n}$, where $X_{a,i}=1$ iff node $a\in\widetilde{G}$ is assigned to $i\in G$.

Having $X$ enables us to map the information to the probability graph. Specifically, the input adjacency matrix is mapped to the predicted adjacency matrix $A' =XAX^T$, the predicted node attributes are $\widetilde{F}'=X^T\widetilde{F}$, and the slices of edge attributes are mapped to $\widetilde{E}'{.,.,l}' =X^T\widetilde{E}'{.,.,l}'X$. The maximum likelihood estimate equals the cross entropy (we assume $F$ and $E$ are one-hot):
\begin{equation}
\begin{array}{c}
\mathcal{L}(\phi, \theta, G) = \mathbb{E}_{q_{\phi}(\boldsymbol{z} \mid G)}\left[-\log p_{\theta}(G \mid \boldsymbol{z})\right]+K L\left[q_{\phi}(\boldsymbol{z} \mid G) \| p(\boldsymbol{z})\right],\\
\log p\left(A^{\prime} \mid \boldsymbol{z}\right) = \frac{1}{k} \sum_{a}\left[A_{a, a}^{\prime} \log \tilde{A_{a, a}}+\left(1-A_{a, a}^{\prime}\right) \log \left(1-A_{a, a}^{\prime}\right)\right]\\+\frac{1}{k(k-1)}
\sum_{a \neq b}\left[A_{a, b}^{\prime} \log A_{a, b}^{\prime}+\left(1-A_{a, b}^{\prime}\right) \log \left(1-A_{a, b}^{\prime}\right)\right], \\
\log p(F \mid \boldsymbol{z}) = \frac{1}{n} \sum_{i} \log F_{i, \cdot}^{T} \tilde{F}_{i, \cdot}^{\prime}, \\
\log p(E \mid \boldsymbol{z}) = \frac{1}{\|A\|_{1}-n} \sum_{i \neq j} \log E_{i, j, r_{i, j,}^{T}}^{T} \tilde{E}_{i,j}^{\prime}.
\end{array}
\end{equation}
The likelihood function considers the existence of matched and unmatched nodes/edges in graphs $\widetilde{G}$ and $G$, but only the attributes of matched nodes and edges. The total reconstruction loss is the weighted sum of three loss terms: the cross-entropy loss for node/edge existence, and the mean squared errorloss for node/edge attributes.
\begin{equation}
-\log p(G \mid \boldsymbol{z})=-\lambda_{A} \log p\left(A^{\prime} \mid \boldsymbol{z}\right)-\lambda_{F} \log p(F \mid \boldsymbol{z})-\lambda_{E} \log p(E \mid \boldsymbol{z})
\end{equation}
\subsection{Metrics of graph similarity}
$\bullet$ Kernel Similarity Index (KSI). It can be used as a metric for comparing graph similarity. It uses graph kernels to calculate the similarity. Graph kernels are feature vectors computed from a graph's adjacency matrix that can be used to measure similarity between graphs.
 The calculation is performed as follows:

$$
KSI\left(G_1,G_2\right)=e^{-\frac{{||E_1-E_2||}^2}{2\sigma^2}},
$$
where ${E}$ is the adjacency matrix of graph, $\sigma = 0.3\times(\left(kernel_{size}-1\right)*0.5-1)+0.8$, where $kernel_{size}$ can be considered the number of nodes contained in the adjacency matrix. A smaller KSI value indicates a higher similarity between the graphs.

$\bullet$  Orbit Count Distribution (Orbit). It measures the number of all orbits with 4 nodes, which reflects higher-level motifs that are shared between generated and raw graphs.
For a graph $G_j$, the sequence for the $i$-th orbit is denoted as $O_i^j =[n_1,n_2...,n_m]$. Then, the calculation is done as follows:

$$
Orbit(G_1,G_2)=\sum_{i=1}^{n}{w_i\ast S i m(O_i^1,O_i^2)},
$$
where $n$ denotes the number of orbits in the graph, $w_i$ is the weight of the $i$-th orbit, $Sim(\cdot)$ represents the similarity calculation function, which can adopt cosine similarity, Jaccard distance, etc. In this work, cosine similarity is used.

$\bullet$ Novel is the fraction of unique correct graphs.

$\bullet$ Unique is the fraction of novel out-of-dataset graphs.

\subsection{Implementation of comparison}


The Fraudulent Transaction Data is a highly imbalanced dataset, with 8,213 fraudulent transaction records and 6,354,407 normal transaction records. In our fraud prediction task, we select all fraudulent transaction records from the original dataset as well as a dataset of normal transactions 10 times the size of the frauds for experimental verification. We consider fraud detection as a binary classification task and use AUC as our main evaluation metric. Additionally, we also count the amount of money involved in the successfully identified fraudulent transactions. All experimental results are the averages of 10 repetitions.


We design two different generative strategies to validate the capabilities of BMS.

$\bullet$  S1: We only introduce the generated structures during training while using real data to validate the BMS's performance of fraud detection.

$\bullet$  S2: We use the original data during training and mix in fraud behaviors based on generated structures during testing to validate the BMS's performance of fraud detection.


%
%
%

\section{Visualization of behavioral structure}
%
%

We select 10 behavior fields (as shown in Table \ref{Table:10fields_visio}) from the Crime Data dataset to create a visualization of the behavioral molecular structure for criminal incidents.
For two behavioral features that are in the form of floating timestamps, we bin them by date and recorded their month and day.
For Figures 1D and 1E, we randomly select 2000 crime incidents from the raw dataset to generate the visualization, where each incident corresponds to ten nodes (representing the values of ten behavioral features) in the visualization. In Figure 1D, we highlight one randomly selected behavior and display its neighbors up to the sixth degree, where lighter colors indicate neighbors that are further away from the highlighted node. In Figure 1E, we count the occurrences of each node among the 2000 incidents, and nodes with larger size and lighter color correspond to higher frequency of occurrence.

\clearpage
\graphicspath{{Figures/}{logo/}}
\clearpage
	\graphicspath{{Figures/}{logo/}}
\clearpage
	\graphicspath{{Figures/}{logo/}}

\clearpage
\graphicspath{{Figures/}{logo/}}

\clearpage
\graphicspath{{Figures/}{logo/}}
\begin{figure*}[!t]
	\centering
	\begin{overpic}[width=0.90\textwidth]{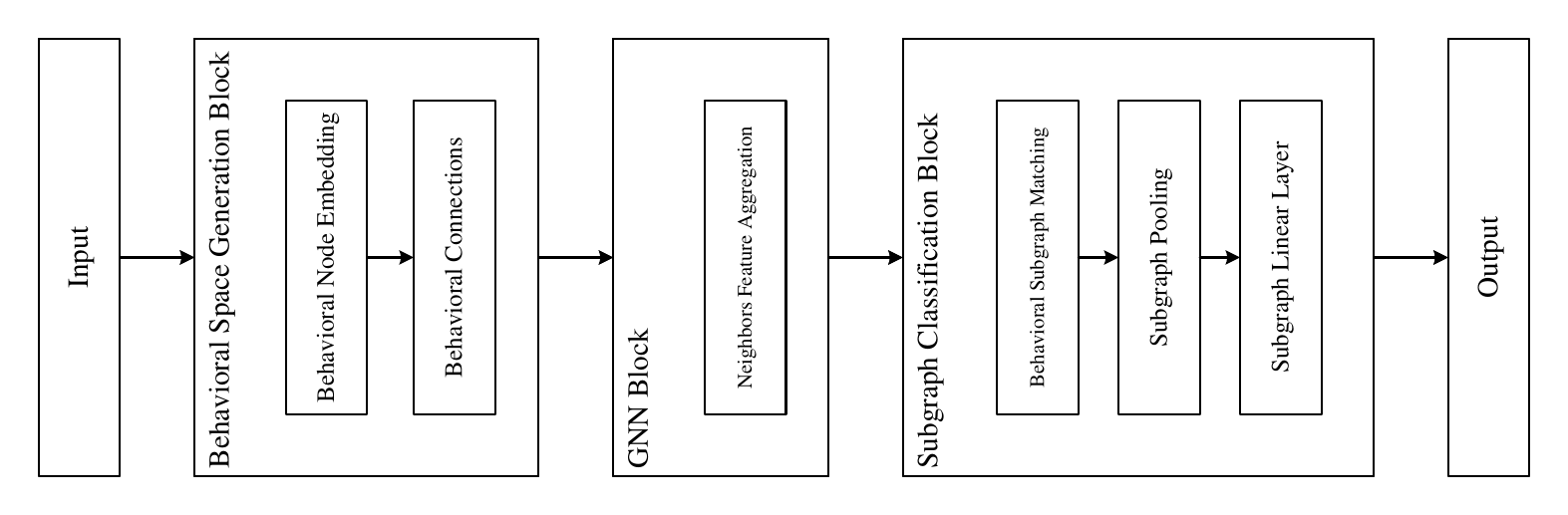}
		\put(-4,27){\large\textbf{A}}
	\end{overpic}
	\begin{overpic}[width=0.90\textwidth]{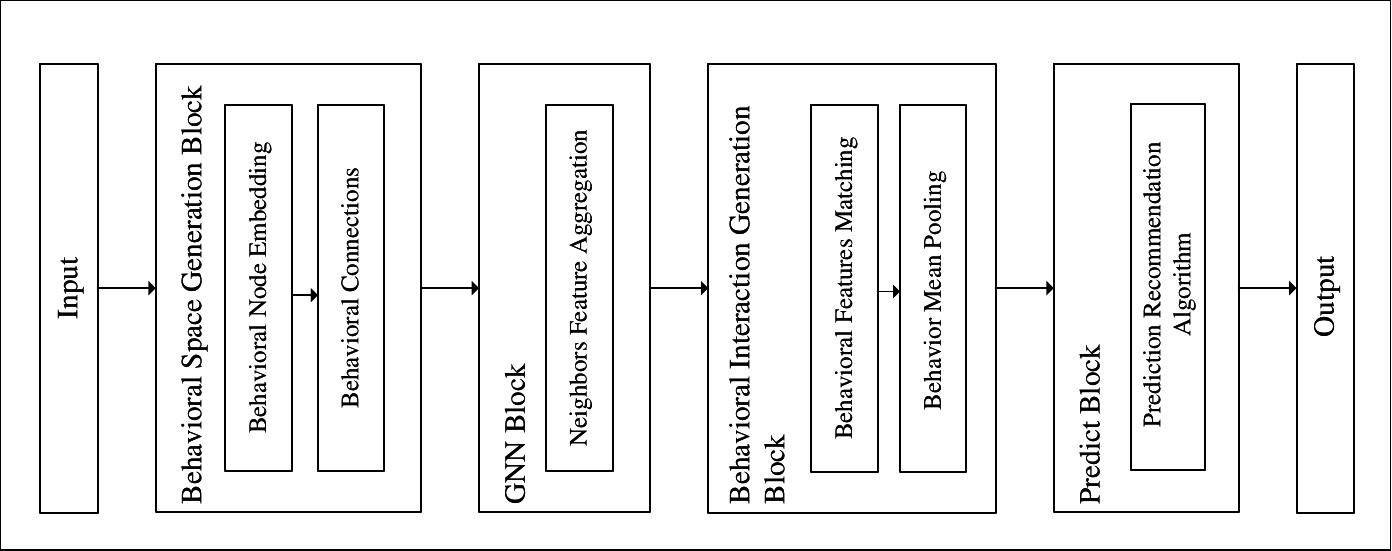}
		\put(-4,34){\large\textbf{B}}
	\end{overpic}
	\begin{overpic}[width=0.90\textwidth]{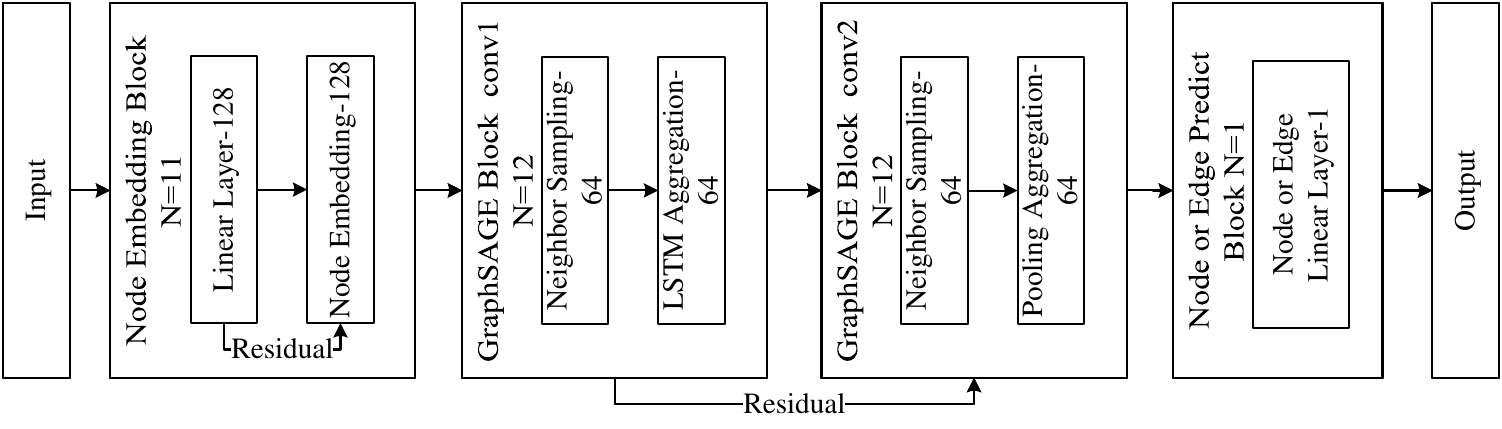}
		\put(-4,25){\large\textbf{C}}
	\end{overpic}

	\captionsetup[subfigure]{labelformat=empty}
	\caption{\textbf{BMS architecture in different downstream tasks.}
		(\textbf{A}) Behavior detection.
		(\textbf{B}) Behavior prediction.
		(\textbf{C}) Behavior generation.
	}
	
	\label{figure:architecture in different downstream tasks}
\end{figure*}

\clearpage

\begin{table}[]
	\centering
	\begin{tabular}{m{5.0cm}<{\raggedright}m{5cm}<{\raggedright} m{5.0cm}<{\raggedright}}
		\toprule
		\textbf{User}                              & \textbf{Answer}                           & \textbf{Impression}                        \\
		\midrule
		user id                           & answer id                        & user id                           \\ \hline
		register timestamp                & question id                      & answer id                         \\ \hline
		gender                            & anonymous or not                 & impression timestamp              \\ \hline
		login frequency                   & author id (null for anonymous)   & click timestamp (0 for non-click) \\ \hline
		\#followers                       & labeled high-value answer or not &                                   \\ \hline
		\#topics followed by this user    & recommended by the editor or not &                                   \\ \hline
		\#questions followed by this user & create timestamp                 &                                   \\ \hline
		\#answers                         & contain pictures or not          &                                   \\ \hline
		\#questions                       & contain videos or not            &                                   \\ \hline
		\#comments                        & \#thanks                         &                                   \\ \hline
		\#thanks received by this user    & \#likes                          &                                   \\ \hline
		\#comments received by this user  & \#comments                       &                                   \\ \hline
		\#likes received by this user     & \#collections                    &                                   \\ \hline
		\#dislikes received by this user  & \#dislikes                       &                                   \\ \hline
		register type                     & \#reports                        &                                   \\ \hline
		register platform                 & \#helpless                       &                                   \\ \hline
		from android or not               & token ids in the answer          &                                   \\ \hline
		from iphone or not                & topic ids of the answer          &                                   \\ \hline
		from ipad or not                  &                                  &                                   \\ \hline
		from pc or not                    &                                  &                                   \\ \hline
		from mobile web or not            &                                  &                                   \\ \hline
		device model                      &                                  &                                   \\ \hline
		device brand                      &                                  &                                   \\ \hline
		platform                          &                                  &                                   \\ \hline
		province                          &                                  &                                   \\ \hline
		city                              &                                  &                                   \\ \hline
		topic ids followed by this user   &                                  &                                  \\
		\bottomrule
	\end{tabular}
	\caption{Fields extracted from the dataset, which include User and Answer information, as well as interaction data Impression (fields with a ``\#'' symbol have numeric feature values).}\label{Table:dataset zhihu}
\end{table}

\clearpage
\begin{table}[]
	\centering
	
	\begin{tabular}{m{3.0cm}<{\raggedright}m{6.3cm}<{\raggedright} m{2.0cm}<{\raggedright}m{3.0cm}<{\raggedright}  }	
		\toprule
		\textbf{Fidelis}        & \textbf{Description}                                                             & \textbf{Type}                               & \textbf{Example}                \\ \midrule
		step           & The time period in which the transaction occurred           & Integer               & 1           \\ \hline
		type           & The type of transaction                                     & String                & PAYMENT   \\ \hline
		amount         & The amount of the transaction                               & Floating Point Value & 9839.64     \\ \hline
		nameOrig       & The account holder who initiated the transaction            & String                & C1231006815 \\ \hline
		oldBalanceOrg  & The account balance of the account holder before the transaction & Floating Point Value & 170136.0 \\ \hline
		newbalanceOrig & The account balance of the account holder after the transaction  & Floating Point Value & 0.0      \\ \hline
		nameDest       & The account holder who received the transaction             & String                & M1979787155 \\ \hline
		oldbalanceDest & The account balance of the recipient before the transaction & Floating Point Value & 0.0         \\ \hline
		newbalanceDest & The account balance of the recipient after the transaction  & Floating Point Value & 0.0         \\ \hline
		isFraud        & A binary indicator of whether the transaction is fraudulent & Integer               & 0 or 1      \\
		\bottomrule
	\end{tabular}
	\caption{Fields and descriptions in the Fraudulent Transaction Data.}\label{Table:dataset fraudulent}
\end{table}


\clearpage
\begin{table}[]
	\centering
	
	\begin{tabular}{m{3.0cm}<{\raggedright}m{6.3cm}<{\raggedright} m{2.0cm}<{\raggedright}m{3.0cm}<{\raggedright}  }	
		\toprule
		\textbf{Fidelis}        & \textbf{Description}                                                             & \textbf{Type}                               & \textbf{Example}                \\ \midrule
		
		Date Rptd      & Report date.                                                            & Floating Timestamp                 & 01/08/2020 12:00:00 AM \\ \hline
		
		DATE OCC       & Date of crime.                                                          & Floating Timestamp                 & 01/08/2020 12:00:00 AM \\\hline
		
		TIME OCC &
		In 24 hour military time. & Plain Text &
		2230 \\\hline
		
		AREA &
		The LAPD has 21 Community Police Stations referred to as Geographic Areas within the department. These Geographic Areas are sequentially numbered from 1-21. &
		Plain Text &
		3 \\\hline
		
		Rpt Dist No    & A four-digit code that represents a sub-area within a Geographic Area.  & Plain Text                         & 377                    \\\hline
		
		Crm Cd         & Indicates the crime committed. (Same as Crime Code 1)                   & Plain Text                         & 624                    \\\hline
		
		Vict Age       & Two character numeric.                                                   & Plain Text                         & 36                     \\\hline
		
		Vict Descent   & Descent Cod.                                                            & Plain Text                         & B                      \\\hline
		
		Premis Cd      & The type of structure, vehicle, or location where the crime took place. & Plain Text                         & 501                    \\\hline
		Weapon Used Cd & The type of weapon used in the crime.                                   & Plain Text & 400                    \\
		\bottomrule
	\end{tabular}
	\caption{Introduction to the selected behavioral features of the Crime Dataset. We provide a criminal incident with case number (DR\_NO 10304468) as an example.}\label{Table:10fields_visio}
\end{table}

\end{document}